\documentclass{revtex4}

\usepackage{amsthm,amssymb,amsmath}				
\usepackage{graphicx,comment}
\usepackage{float}   
\usepackage{xcolor}

\begin{document}

\title{PDM Klein-Gordon oscillators in cosmic string spacetime in magnetic and Aharonov-Bohm flux fields within the Kaluza-Klein theory}
\author{Omar Mustafa}
\email{omar.mustafa@emu.edu.tr}
\affiliation{Department of Physics, Eastern Mediterranean University, G. Magusa, North
Cyprus, Mersin 10 - Turkey.}

\begin{abstract}
\textbf{Abstract:}\ In the cosmic string spacetime and within Kaluza-Klein theory (KKT) backgrounds (indulging magnetic and Aharonov-Bohm flux fields), we introduce and study position-dependent mass (PDM) Klein-Gordon (KG) oscillators.  The effective PDM is introduced as a deformation/defect in the momentum operator. We show that there are four different ways to obtain KG-oscillator. Two of which are readily known and the other two are obtained as byproducts of PDM settings. Next, we provide a thorough analysis on the corresponding spectra under different parametric effects, including the curvature parameter's effect. Such analysis is used as a reference/lead model which is used in the discussion of different PDM KG-oscillators models: a  mixed power-law and exponential type  PDM model that yields a pseudo-confined PDM KG-oscillator in cosmic string spacetime within KKT (i.e., the PDM KG-oscillators are confined in their own PDM manifested Cornell-type confinement), and a PDM KG-oscillator confined in a Cornell-type potential. Moreover, we extend our study and discuss a non-Hermitian $\mathcal{PT}$-symmetric PDM-Coulombic-type KG-particle model in cosmic string spacetime within KKT.


\end{abstract}

\maketitle

\section{Introduction}
Topological defects as products of disclination, dislocation, and dispiration in media or in cosmic string spacetime manifestly affect the physical properties of the system at hand \cite{R1,R2}. They appear as monopole, strings, and walls in gravitation \cite{R3}, as vortices in superconductors or superfuilds, domain walls in magnetic materials, solitons in polymers, and disclinations or dislocations in solids and liquid crystals \cite{R4,R5,Geusa 2001}. Studies of quantum mechanical relativistic particles in spacetime with cosmic strings as topological defects are ample in the literature. It has been shown, for example, that the energy spectrum in a curved spacetime is different than that in the flat Minkowski spacetime \cite{R7,R8,R9,R10,R11,Ahmed 1,Ahmed 2,Santos 2018,R12}. 
The Dirac and Klein-Gordon particles in a G\"{o}del-type spacetime background at positive, negative , and zero curvatures are studied in \cite{Figueiredo 1992}. Nevertheless, the Dirac oscillator of Moshinsky and Szczepaniak \cite{Moshinsky 1989} has encouraged research activities on Klein-Gordon (KG) oscillator \cite{Bruce 1993,Dvoeg 1994}. For example, the KG-oscillator in the G\"{o}del and G\"{o}del-type spacetime backgrounds (e.g., \cite {Moshinsky 1989,Bruce 1993,Dvoeg 1994,Das 2008,Carvalho 2016,Garcia 2017,Vitoria 2016,Vitoria1 2016,Ahmed1 2018,Lutfuoglu 2020}), in cosmic string spacetime within Kaluza-Klein theory (KKT) backgrounds (e.g., \cite{Ahmed 2,Ahmed1 2018,Ahmed1 2020,Ahmed1 2021,Boumal 2014,R13}), in Som-Raychaudhuri \cite{Wang 2015}, in the (2+1)-dimensional G\"{u}rses spacetime backgrounds (e.g., \cite{Gurses 1994,Ahmed 2019,Ahmed1 2019,Ahmed2 2019}), etc. 

Position-dependent mass (PDM), on the other hand, have attracted researchers attention over the years after the introduction of the Mathews-Lakshmanan oscillator \cite{M-L 1974}, both in classical and quantum mechanics \cite{M-L 1974,von Roos,Carinena Ranada Sant 2004,Mustafa 2019,Mustafa1 2020,Mustafa arXiv,Mustafa Phys.Scr. 2020,Mustafa Habib 2007,Mustafa Algadhi 2019,Khlevniuk 2018,Mustafa 2015,Dutra Almeida 2000,dos Santos 2021,Nabulsi1 2020,Nabulsi2 2020,Nabulsi3 2021,Quesne 2015,Tiwari 2013}. Such a PDM concept does not necessarily mean that the mass is literally position dependent. PDM means that the mass becomes effectively position-dependent. It has been shown that the PDM concept is, in fact, a metaphoric manifestation of coordinate
deformation/transformation \cite{Mustafa1 2020,Mustafa arXiv,Mustafa Phys.Scr. 2020,Khlevniuk 2018}. The coordinate deformation/transformation, in effect, changes the form of the canonical momentum in classical and the momentum operator in quantum mechanics (e.g., \cite{Mustafa1 2020,Mustafa arXiv,Mustafa Algadhi 2019,dos Santos 2021} and related references therein). In classical mechanics, for example, negative the gradient of the potential force field is shown to be related to the time derivative of the pseudo-momentum (also called Noether momentum) $\pi \left( x\right) =\sqrt{%
m\left( x\right) }\dot{x}$ \cite{Mustafa arXiv}, and is no longer given by the time derivative of the canonical momentum $p=m\left(
x\right) \dot{x}$.  In quantum mechanics, however, the PDM momentum operator is constructed by Mustafa and Algadhi \cite{Mustafa Algadhi 2019} and is used to find the PDM creation and annihilation operators for the\ PDM-Schr\"{o}dinger oscillator \cite{Mustafa1 2020}.  It would be interesting, therefore, to investigate the effects of PDM on the KG-oscillator (in particular, and KG-particles in general) in cosmic string spacetime background within KKT.

Attempts were, in fact, made to include PDM settings in the Dirac and/or KG relativistic equations \cite{Mustafa Habib 2008,Mustafa Habib1 2007,Vitoria
Bakke 2016} through the assumption that $m\longrightarrow m+m\left( r\right)
+S\left( r\right) =M\left( r\right) $, where $m$ denotes the rest mass energy, $%
m\left( r\right) $ PDM, $S\left( r\right) $ the Lorentz scalar potential interaction, and $M\left( r\right) $ denotes effective PDM in the relativistic wave equation at hand. However, in the current methodical proposal, we stay away from such PDM perception and adopt the fact that PDM is a manifestation of coordinate deformation/transformation and, therefore, use the
PDM-quantum mechanical operator \cite{Mustafa Algadhi 2019,Mustafa1 2020,dos Santos 2021} (see (\ref{a5}) below). Under such settings, we shall study
PDM KG-oscillator in cosmic string spacetime within KKT.

The organization of our methodical proposal is in order. In section 2, we introduce the PDM KG-oscillator in cosmic string spacetime within KKT. Where the effective PDM is introduced as a defect/deformation in the momentum operator. Moreover, we show that there are four different ways to obtain KG-oscillators. Two of which are readily known and the other two (new) are byproducts of PDM settings. In section 3, we recycle/recollect the KG-oscillators in cosmic string spacetime in magnetic and Aharonov-Bohm flux fields within the KKT, and provide a thorough analysis on the corresponding spectra under different parametric effects. In so doing, we provide a reference/lead model that is to be used in the following sections on different PDM KG-oscillators models in cosmic string spacetime within KKT.  In section 4, we discuss a mixed power-law and exponential type  PDM model that yields a pseudo-confined PDM KG-oscillator in cosmic string spacetime within KKT. This confinement is introduced as a byproduct of the PDM settings of the KG-oscillator (i.e., the PDM KG-oscillators are confined in their own PDM manifested confinement), hence the notion \textit{"pseudo-confined PDM KG-oscillators"}. In section 5, we discuss a confined PDM KG-oscillator-III (the third of the four KG-oscillators reported in section 2) in cosmic string spacetime within KKT. In section 6, we use a $\mathcal{PT}$-symmetric PDM-Coulombic-type KG-particle model in cosmic string spacetime within KKT. Hereby, we follow the Bender and Boettcher \cite{Bender 1998} relaxation of the Hermiticity condition on quantum Hamiltonians that  have resulted in the introduction of the nowadays called $\mathcal{PT}$-symmetric quantum mechanics, which yields real spectra (for more details on this issue the reader may refer to
\cite{Znojil Levai 2000,Mustafa Znojil 2002} and references cited therein). Where, $\mathcal{P}$ denotes space reflection: $x\rightarrow-x$, $\mathcal{T}$ mimics the time-reversal: $i\rightarrow-i$, and a $\mathcal{PT}$-symmetric function would satisfy $f(x)=|f(-x)|^{*}$.  We give our concluding remarks in section 7.

\section{PDM KG-oscillators in cosmic string spacetime in magnetic and Aharonov-Bohm flux fields within KKT}

The Kaluza-Klein theory proposes an extra compact space dimension while introducing pure gravity in the new $(4+1)$-dimensional spacetime. The idealized assumption in the KKT is that the topological defect carries in its core a magnetic field with a magnetic flux $\Phi$ (i.e., Aharonov-Bohm flux field) that vanish outside the defect \cite{Geusa 2001,Furtado 2000}. Under such assumption, the magnetic field is not introduced via minimal coupling but is rather treated as a topological defect within the framework of the Kaluza-Klein theory. The corresponding cosmic string spacetime metric (in $G=\hbar=c=1$ units) is given by%
\begin{equation}
ds^{2}=-dt^2+dr^2+\alpha^2\,r^2\,d\varphi^2+dz^2+(du-\kappa\,A_{\mu}\,dx^{\mu})^2,
\label{a1}
\end{equation}%
where the fifth dimension $u$ is a space-like and it varies as $0 < u < 2\,\pi\,a$, where $a$ is the radius of the compact dimension of $u$. Here, $\kappa $ is the Kaluza constant, and $\alpha =( 1-4\,\tilde{\mu}) $ with the linear mass density of the string $\tilde{\mu}$ \cite{Santos 2018}. The parameter $\alpha$ assumes values $0<\alpha<1$ for positive curvature (in general relativity for cosmic string with positive mass density), $1<\alpha<\infty$ for negative curvature (in the geometric theory of defects in condensed matter), and $\alpha=1$ corresponds to Minkowski flat spacetime. Moreover, the coordinates ranges are $\varphi\in[0,2\pi]$, $r\in[0,\infty]$, and $(z,t)\in[-\infty,\infty]$. We now assume that the electromagnetic four-vector potential, indulging the Aharonov-Bohm flux field $\Phi$, within KKT is given by%
\begin{equation}
A_{\mu}=(0, 0, A_{\varphi}, 0)\,;\,\, A_{\varphi}=\kappa^{-1}\,\left[\frac{1}{
2\,\alpha}\,B_{\circ }\,r+\frac{\Phi}{2\,\pi\,\alpha\,r}\right],
\label{a2-1}
\end{equation}%
so that%
\begin{equation}
\kappa\,A_{\mu}\,dx^{\mu}=\kappa\,A_{\varphi}\,\alpha\,r\,d\varphi=
\left(\frac{1}{2}\,B_{\circ }\,r^2+\frac{\Phi}{2\,\pi}\right)\,d\varphi,
\label{a2-2}
\end{equation}%
where a uniform magnetic field is manifested in the $z$-direction through $\mathbf{B=\nabla \times A}=(\kappa^{-1}B_{\circ }/\alpha)\,\hat{z}$, and the Aharonov-Bohm flux field $\Phi$ is indulged in the second term of $A_{\varphi}$  \cite{Ahmed 1,Geusa 2001,Furtado 2000,Mustafa Z 2020}. Here, we have used $\mathbf{A}=\mathbf{A}_{1}+\mathbf{A}_{2}$  so that $
\mathbf{A}_{1}=\left(0, \frac{\kappa^{-1}}{2\,\alpha}\,B_{\circ }\,r, 0\right)$, and $\mathbf{A}
_{2}=\left(0,\frac{\kappa^{-1}\,\Phi}{2\,\pi\,\alpha r}, 0\right)$, with $\mathbf{\nabla \times A}
_{1}=(\kappa^{-1}B_{\circ }/\alpha)\,\hat{z}$, and $\mathbf{\nabla \times A}_{2}=0$. Hence, the cosmic string in $(4+1)$-dimensional spacetime metric within KKT reads%
\begin{equation}
ds^2=-dt^2+dr^2+\alpha^2\,r^2\,d\varphi^2+dz^2+\left[du+\tilde{A}_{\varphi}\,d\varphi \right]^2;\,\tilde{A}_{\varphi}=-\left(
\frac{1}{2}\,B_{\circ }\,r^2+\frac{\Phi}{2\,\pi}\right).
\label{a3}
\end{equation}%
Then the corresponding covariant and contravariant metric tensors in this case, respectively, read
\begin{equation}
g_{\mu \nu }=\left( 
\begin{tabular}{ccccc}
$-1\smallskip $ & $\,0\,$ & $0$ & $\,0$ & $0$ \\ 
$0$ & $1\smallskip $ & $0$ & $0$ & $0$ \\ 
$0$ & $\,0$ & $\,\left(\alpha^2\,r^2+\tilde{A}_{\varphi}^2\right) \,$
& $0$ & $\tilde{A}_{\varphi}$ \\ 
$0$ & $0$ & $0$ & $1$ & $0$ \\ 
$0$ & $0$ & $\tilde{A}_{\varphi}$ & $0$ & $1$%
\end{tabular}%
\right) \Leftrightarrow g^{\mu\nu}=\left( 
\begin{tabular}{ccccc}
$-1\smallskip $ & $\,0\,$ & $0$ & $0$ & $0$ \\ 
$0$ & $\,1\smallskip $ & $0$ & $0$ & $0$ \\ 
$0$ & $\,0$ & $\,\frac{1}{\alpha^2\,r^2}\,$ & $-0$ & $-\frac{\tilde{A}_{\varphi}}{\alpha ^2\,r^2}$ \\ 
$0$ & $0$ & $-0\,$ & $\,1$ & $0$ \\ 
$0$ & $0$ & $-\frac{\tilde{A}_{\varphi}}{\alpha^2\,r^2}$ & $0$ & $\left(
1+\frac{\tilde{A}_{\varphi}^2}{\alpha^2\,r^2}\right) $%
\end{tabular}%
\right)
\label{a4}
\end{equation}%
where $\det \left( g\right) =-\alpha^2\,r^2.$ 

Very recently, however, Mustafa and Algadhi \cite{Mustafa Algadhi 2019} have introduced the PDM-momentum operator%
\begin{equation}
\mathbf{\hat{p}}\left( \mathbf{r}\right) =-i\,\left( \mathbf{\nabla -}\frac{%
\mathbf{\nabla }m\left( \mathbf{r}\right) }{4\,m\left( \mathbf{r}\right) }%
\right) \Longleftrightarrow p_{j}=-i\,\left( \partial_{j}-\frac{\partial_{j}\,m\left( \mathbf{r}\right) }{4\,m\left( \mathbf{r}\right) }\right) ,
\label{a5}
\end{equation}%
which is used to study the torsion effects on the PDM Klein-Gordon oscillator in a cosmic string spacetime background \cite{Mustafa arXiv1}. In the current study, we shall use such PDM-momentum operator \cite{Mustafa Algadhi 2019} and consider a PDM KG-oscillator with a uniform magnetic field and Aharonov-Bohm flux field within the KKT. Moreover, we shall use the assumption that $m(\textbf{r})=m(r)$ (i.e., the PDM function is only radially dependent). Under such settings, the momentum operator would take the PDM form \cite{Mustafa arXiv1} so that%
\begin{equation}
\tilde{p}_{\mu}\longrightarrow -i\,\partial_{\mu }+i\,\mathcal{F}_{\mu }%
\,{;}\quad \mathcal{F}_{\mu }=\left(0,\eta\,r+\frac{m^{\prime}\left(
r\right)}{4\,m\left( r\right) }, 0, 0\right) \Longrightarrow \mathcal{F}%
_{r}=\eta\,r+\frac{m^{\prime}\left(r\right)}{4\,m\left(r\right)},
\label{a6}
\end{equation}%
which is to be used to construct the PDM KG-oscillator in the presence of a Lorentz scalar potential $S(r)$ as follows%
\begin{equation}
\frac{1}{\sqrt{-g}}\,\left(\partial_{\mu}+\mathcal{F}_{\mu }\right)\,\left[ 
\sqrt{-g}\,g^{\mu\nu}\,\left(\partial_{\nu }-\mathcal{F}_{\nu}\right)\,\Psi %
\right] =(m + S(r))^2\,\Psi .  
\label{a7}
\end{equation}%
This would result, with $%
\Psi =e^{i\,(-E\,t+l\,\varphi+k_{z}\,z+Q\,u)}\,R\left( r\right) $, in%
\begin{equation}
R^{\prime \prime }\left( r\right) +\frac{1}{r}R^{\prime }\left( r\right) +%
\left[ \lambda -\frac{\tilde{\gamma}^{2}}{r^{2}}-\omega ^{2}r^{2}-\left( 
\mathcal{F}_{r}^{\prime }+\frac{\mathcal{F}_{r}}{r}+\mathcal{F}%
_{r}^{2}\right) -2mS\left( r\right) -S\left( r\right) ^{2}\right] R\left(
r\right) =0,
\label{a8}
\end{equation}%
and the substitution $R\left( r\right) =U\left( r\right) /\sqrt{r}$ would imply%
\begin{equation}
U^{\prime \prime }\left( r\right) +\left[ \lambda -\frac{\left( \tilde{\gamma%
}^{2}-1/4\right) }{r^{2}}-\omega ^{2}r^{2}-\left( \mathcal{F}_{r}^{\prime }+%
\frac{\mathcal{F}_{r}}{r}+\mathcal{F}_{r}^{2}\right) -2mS\left( r\right)
-S\left( r\right) ^{2}\right] U\left( r\right) =0,
\label{a9}
\end{equation}%
where%
\begin{equation}
\lambda =E^{2}-Q^{2}-k_{z}^{2}-m^{2}-2\omega%
\tilde{\gamma}\,;\,\,\tilde{\gamma}^{2}=\left( \frac{\ell }{\alpha }+\frac{%
 \Phi Q}{2\,\alpha\,\pi  }\right) ^{2}\, ;\,\,\omega =\frac{ QB_{\circ }}{2\alpha }.
\label{a10}
\end{equation}%
Eventually, with $\mathcal{F}_{r}$ of (\ref{a6}), we obtain%
\begin{equation}
U^{\prime \prime }\left(r\right) +\left[\tilde{\lambda}-\frac{\left( 
\tilde{\gamma}^{2}-1/4\right)}{r^2}-\tilde{\omega}^2\,r^2-2\,m\,S\left(
r\right)-S\left(r\right)^2+M\left( r\right) \right]\,U\left( r\right) =0,
\label{a11}
\end{equation}%
where $\tilde{\lambda}=\lambda-2\,\eta$,  $\tilde{\omega}^2=\omega^2+\eta^2$, and 
\begin{equation}
M\left(r\right)=\frac{3}{16}\,\left( \frac{m^{\prime }\left(r\right)}{%
m\left(r\right)}\right)^2-\frac{1}{4}\,\frac{m^{\prime\prime }\left(
r\right)}{m\left(r\right)}-\frac{m^{\prime }\left(r\right)}{4\,r\,m\left(
r\right)}-\eta\,r\,\frac{m^{\prime }\left(r\right)}{2\,m\left(r\right)}.
\label{a12}
\end{equation}%
Under the current proposal settings, it could be interesting to report that there are four feasible models that yield KG-oscillators:
\begin{enumerate}
    \item KG-oscillator-I: is obtained by setting $\mathcal{F}_{r}=\eta\,r$, $B_{\circ }=0=\Phi$ (i.e., no magnetic fields), $S(r)=0$ (i.e., no confinement) and  $m(r)=1$ (i.e., constant mass settings as in Mirza et al's recipe \cite{Mirza 2004}. This would result in the KG-oscillator with $V_{eff}(r)=\eta^2\,r^2+2\,\eta$.%
    \item KG-Oscillator-II: is obtained by setting  $\eta=0$, $m(r)=1$, $S(r)=0$ and $B_{\circ}\neq0$, which implies a KG-oscillator with $V_{eff}(r)=\omega^2\,r^2$,%
     \item KG-oscillator-III: is obtained by setting $\eta=0=B_{\circ}$, $S(r)=0$ and an effective PDM in the form of 
               \begin{equation}
    m(r)=A\,e^{2\,\Omega\,r^2},
    \label{a12-0}
    \end{equation}that yields a shifted by a constant PDM KG-oscillator with $V_{eff}(r)=-M(r)=\Omega^2\,r^2+2\,\Omega$. However, one should notice the resemblance between this KG-oscillator-III and the KG-oscillator-I above. Consequently, the KG- oscillator-I of Mirza et al's recipe \cite{Mirza 2004} or KG-oscillator-III may very well be considered as  byproducts of PDM settings %
    \item KG-oscillator-IV: is obtained by setting $\eta=0=B_{\circ}$,  $S(r)=0$ and an effective PDM of complex settings in the form of%
    \begin{equation}
    m \left( r \right) =C\,{\frac { {{ I}_{0}\left(\frac{\Omega\,{r}^{2}}{2}\right)}^{4}}{{\Omega}^{4}{r}^{8} \left[ {{ I}_{0}\left(\frac{\Omega\,{r}^{2}}{2}\right)}{{K}_{1}\left(\frac{\Omega\,{r}^{2}}{2}\right)}+{{ I}_{1}\left(\frac{\Omega\,{r}^{2}}{2}\right)}{{ K}_{0}\left(\frac{\Omega\,{r}^{2}}{2}\right)} \right] ^{4}}}, \label{a12-1}
    \end{equation}where $I_{i}(x)$ and $K_i(x)$ are the modified Bessel functions. would result $M(r)=-\Omega^2\,r^2$, in (\ref{a12}), and consequently yielding a PDM KG-oscillator with $V_{eff}(r)=\Omega^2\,r^2$.%
\end{enumerate}%
Notably, one may figure out that the KG-oscillator-I by  Mirza's et al recipe \cite{Mirza 2004} turns out to be nothings but a byproduct of the PDM setting in KG-oscillator-III of (\ref{a12-0}). Effectively, the topological defect in the momentum operator suggested by Mirza et al. may very well be explained as an effective topological defect of the effective PDM settings. That is, if one sets $\eta=0$ (i.e., dismissing Mirza's term) then the effective frequency of the emerging PDM KG-oscillator is now given by $\Omega$. Consequently, the current methodical proposal may mark a new and more general PDM KG-particles (not only KG-oscillators) in cosmic string spacetime or in the geometric theory of topological defects in condensed matter.

In the forthcoming sections, we start with the most simplistic KG-oscillator that would serve as a reference model to check on the exact solvability of the more general and more complicated models to be discussed  in the sequential sections. Yet, the four KG-oscillator models above are exactly solvable within the lines discussed in the following section.

\section{KG-oscillator in cosmic string spacetime within KKT: recycled and analyzed}

In this section, we assume the Lorentz scalar potential $S(r)=0$ and the PDM function $m\left( r\right)=1$ (i.e., constant mass settings). Therefore, equation (\ref{a11}) becomes %
\begin{equation}
U^{\prime\prime} (r)+\left[\tilde{\lambda}-\frac{\left( 
\tilde{\gamma}^{2}-1/4\right)}{r^2}-\tilde{\omega}^2\,r^2\right]\,U (r) =0,
\label{a13}
\end{equation}%
where  $\tilde{\lambda}=\lambda-2\,\eta$,  $\tilde{\omega}^2=\omega^2+\eta^2$ (this KG-oscillator covers both KG-oscillator-I and/or II, above). Moreover, this equation resembles the two-dimensional radial Schr\"{o}dinger harmonic oscillator with an effective oscillation frequency $\tilde{\omega}$ and an irrational magnetic quantum number $\tilde\gamma$. Consequently, it inherits its textbook eigenvalues%
\begin{equation}
\tilde{\lambda}=2\,\tilde{\omega}\,\left(2\,n_{r}+\left\vert \tilde{\gamma}%
\right\vert +1\right)\Longrightarrow\lambda-2\,\eta =2\,\tilde{\omega}\,\left(2\,n_{r}+\left\vert \tilde{\gamma}%
\right\vert +1\right),
\label{14-1}
\end{equation}%
where $n_r=0,1,2,\cdots$ is the radial quantum number (denotes number of nodes in the wave function) (c.f., e.g., \cite{Furtado 2000,Bezerra 2019,Bakke 2010,Mota 2014}). Hence, the KG-oscillator of (\ref{a13}) admits exact energy levels given by%
\begin{equation}
E^2_{n_{r},l}=2\,\tilde{\omega}\,\left(
2\,n_{r}+\left\vert \frac{\ell }{\alpha}+\frac{\,\Phi\,Q}{2\,\alpha\,\pi}%
\right\vert +1\right) +\frac{Q\,B_{\circ}}{\alpha}\left( \frac{\ell }{%
\alpha }+\frac{\Phi\,Q}{2\,\alpha\,\pi }\right)
+k_{z}^2+m^2+Q^2+2\,\eta,
\label{a14-2}
\end{equation}%
and the corresponding unnormalized radial eigenfunctions are given by%
\begin{equation}
U\left( r\right) \sim r^{\left\vert \tilde{\gamma}\right\vert +1/2}\exp
\left( -\frac{\tilde{\omega}r^{2}}{2}\right) L_{n_{r}}^{\left\vert \tilde{%
\gamma}\right\vert }\left( \tilde{\omega}r^{2}\right) \Longleftrightarrow
R\left( r\right) \sim r^{\left\vert \tilde{\gamma}\right\vert }\exp \left( -%
\frac{\tilde{\omega}r^{2}}{2}\right) L_{n_{r}}^{\left\vert \tilde{\gamma}%
\right\vert }\left( \tilde{\omega}r^{2}\right),
\label{a15}
\end{equation}%
\begin{figure}[!ht]  
\centering
\includegraphics[width=0.3\textwidth]{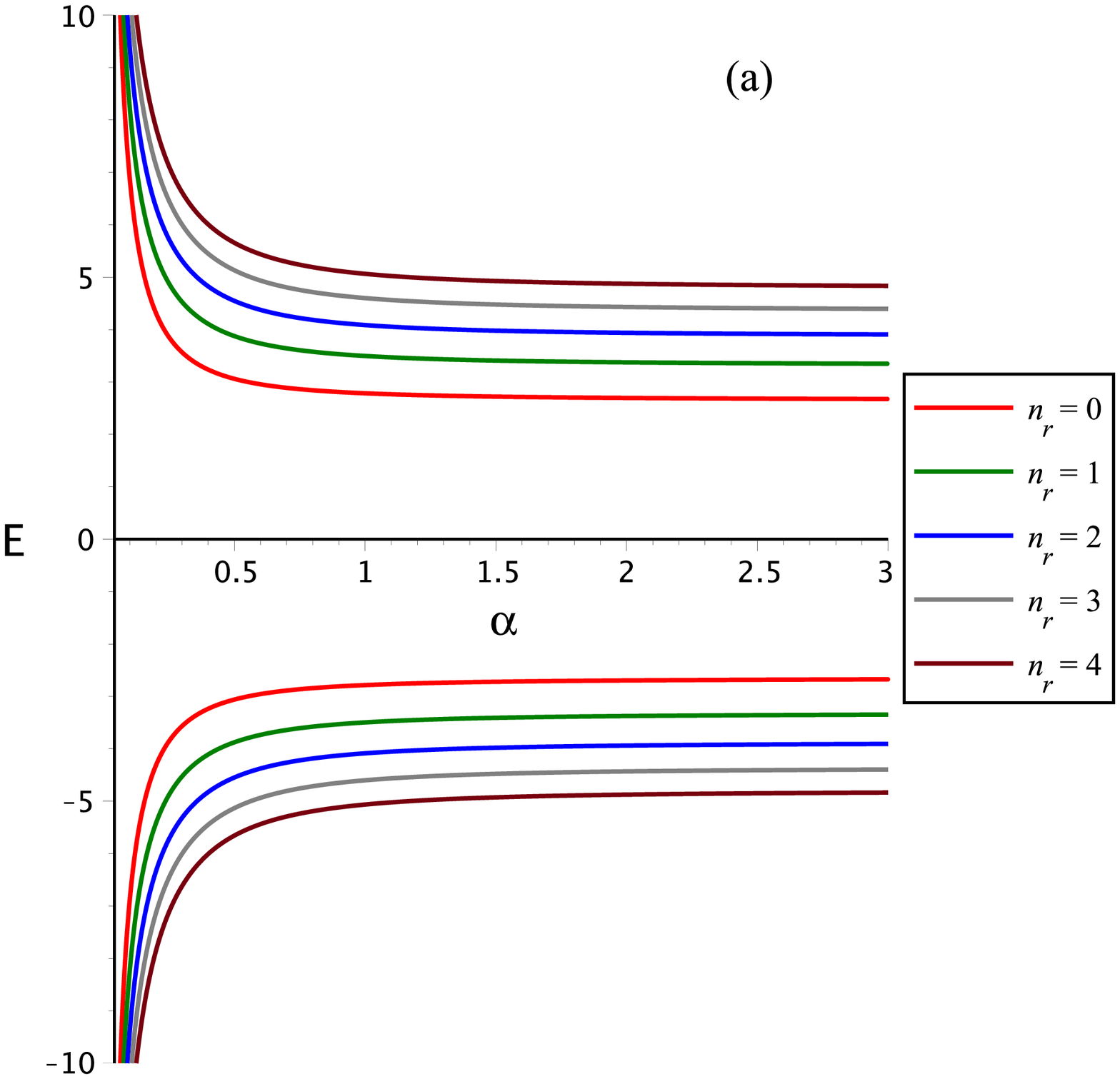}
\includegraphics[width=0.3\textwidth]{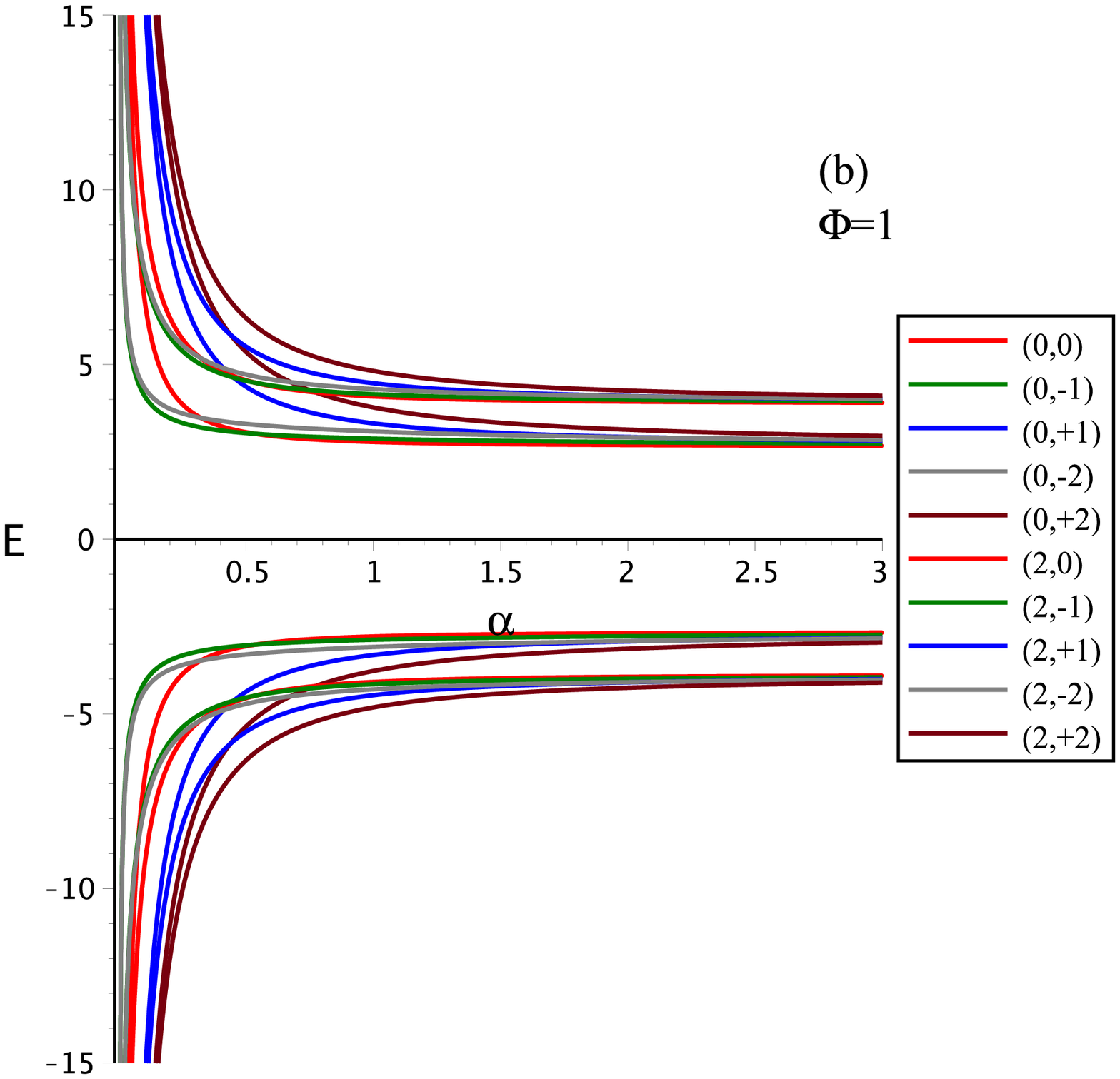}
\includegraphics[width=0.3\textwidth]{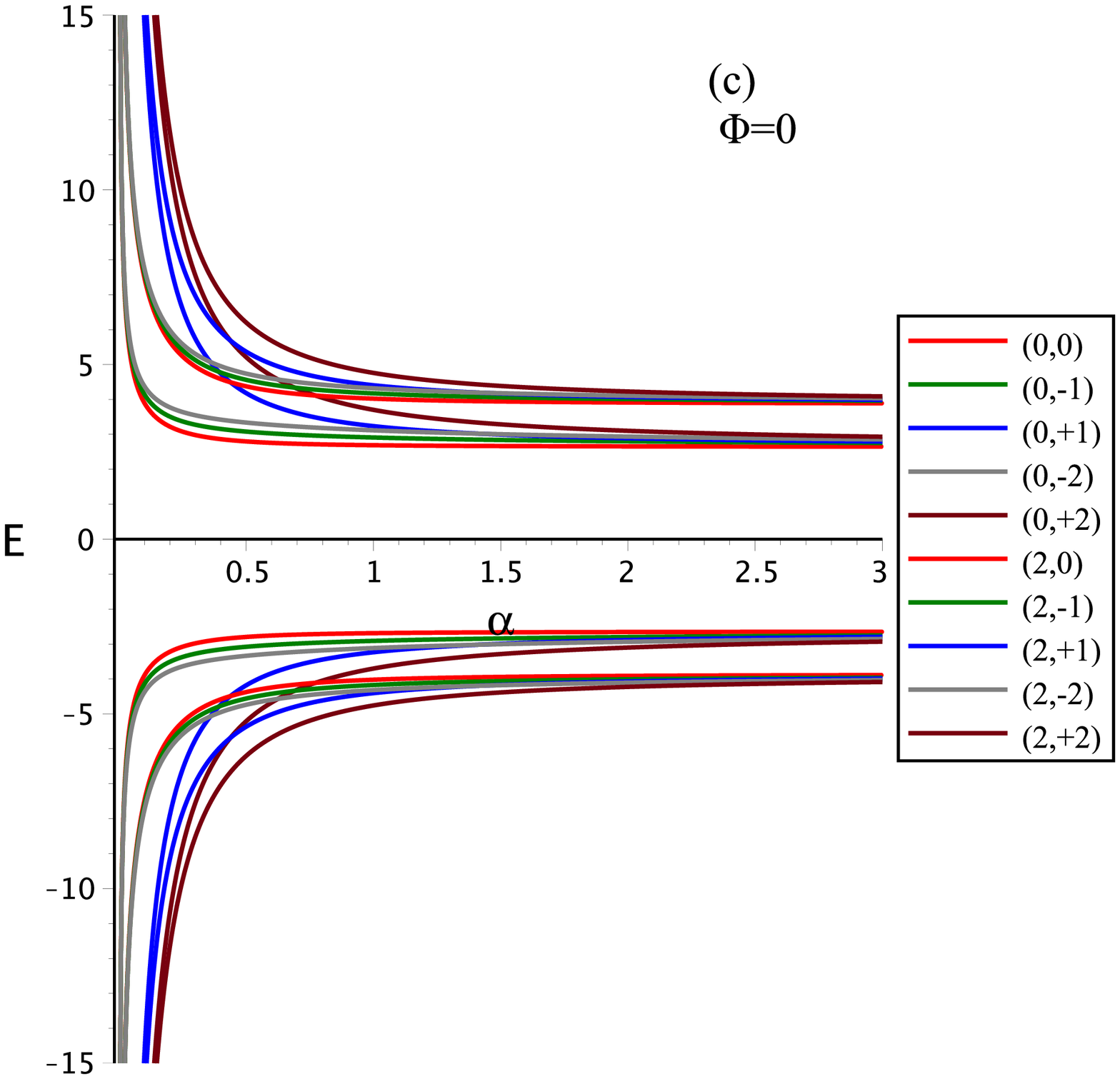}
\caption{\small 
{ For $m=k_z=Q=\Phi=B_{\circ}=\eta=1$ we plot the energy levels $E_{n_r,\ell}$  so that in (a) $E_{n_r,0}$ at $0<\alpha<10$ for $n_r=0,1,2,3,4$, and (b) $E_{n_r,\ell}$ at $0<\alpha<3$ for $n_r=0,2$ and $\ell=0,\pm1,\pm2$. For $m=k_z=Q=B_{\circ}=\eta=1$,  $\Phi=0$ we plot in (c) $E_{n_r,\ell}$ at  $0<\alpha<3$ for $n_r=0,2$ and $\ell=0,\pm1,\pm2$.}}
\label{fig1}
\end{figure}%
where $ L_{n_{r}}^{\left\vert \tilde{\gamma}\right\vert }\left( \tilde{\omega}r^{2}\right)$ are the associated Laguerre polynomials that are, up to a multiplicity constant, related to the confluent hypergeometric polynomials $_{1}F_{1}(-n_r,1+|\tilde\gamma|,\tilde{\omega}\,r^2)$. Nevertheless, one should notice that for $S(r)\neq 0$ and/or $M(r)\neq 0$, the solution of such a more general problem should collapse back to (\ref{a14-2}) and (\ref{a15}) for the case $S(r)= 0$ and $M(r)=0$.  Moreover, it is obvious that square the energies in (\ref{a14-2}) (i.e., $E^2$) resembles Landau-type energies-squared with an irrational magnetic quantum number $\tilde{\gamma}$ that indulges within the Aharonov-Bohm flux field effect. It is interesting to know that this is manifestly introduced by the coupling between the compact Kaluza-Klein fifth-dimension $u$ and the $\varphi$ coordinate.  As such, we may argue that the additional compact dimension of the Kaluza-Klein theory offers quasi-Landau energy levels (e.g.,  \cite{Geusa 2001,Bezerra 2019,Zeinab 2020}). Moreover, one should be aware that the four possible KG-oscillators' settings discussed above, with some straightforward parametric substitutions, admit exact solvability as in (\ref{a14-2}) and (\ref{a15}).

In Fig. 1 we plot the energy levels $E_{n_r,\ell}$  against the parameter $\alpha$. In Fig.1(a) we observe that as $\alpha$ increases within the range $0<\alpha<1$ (i.e., cosmic string with positive mass density), the energy gap shrinks and the energy levels separate apart from clustering near $\alpha\approx0$. Whereas, as $\alpha$ increases within the range $1<\alpha<\infty$, the energy gap stabilizes and the energy levels $E_{n_r,0}$ maintain constant (but not equal) spacing as $\alpha$ grows up from 1. Fig. 1(b), shows that for a given $n_r$ the energy  levels $E_{n_r,\ell}$  would cluster and batch up (without levels crossings) around the corresponding $E_{n_r,0}$ energy level (i.e., corresponding S-state) for $\alpha>>1$ within the range $1<\alpha<\infty$. Whereas, a different trend of batching up (clustering) is obvious as $\alpha\rightarrow0$. That is, state of the same magnetic quantum number $\ell$ (i.e., states with the same color) batch up and cluster together. Two different trends of clustering, as $\alpha\rightarrow0$ and as $\alpha\rightarrow\infty$, would necessarily imply that energy levels crossings are eminent and unavoidable in the process. In Fig. 1(c), we set the magnetic flux field $\Phi=0$ and found that the batching-up phenomenon has nothings to do with the Aharonov-Bohm flux field $\Phi$, but it is rather a characterization of the geometric theory of topological defects in condensed matter as well as in cosmic string spacetime, as $\alpha\rightarrow0$ and as $\alpha\rightarrow\infty$ from $\alpha=1$, respectively. Notably, at $\alpha=1$ (i.e., Minkowski flat spacetime) the energy levels maintain their regular quasi-Landau-type structure. Mathematically speaking, two terms in the energy equation (\ref{a14-2}), the first and the second terms, determine the fate of the energy levels. If we temporarily set $\eta=0$ in (\ref{a14-2})  we get the first two terms as%
\begin{figure}[!ht]  
\centering
\includegraphics[width=0.3\textwidth]{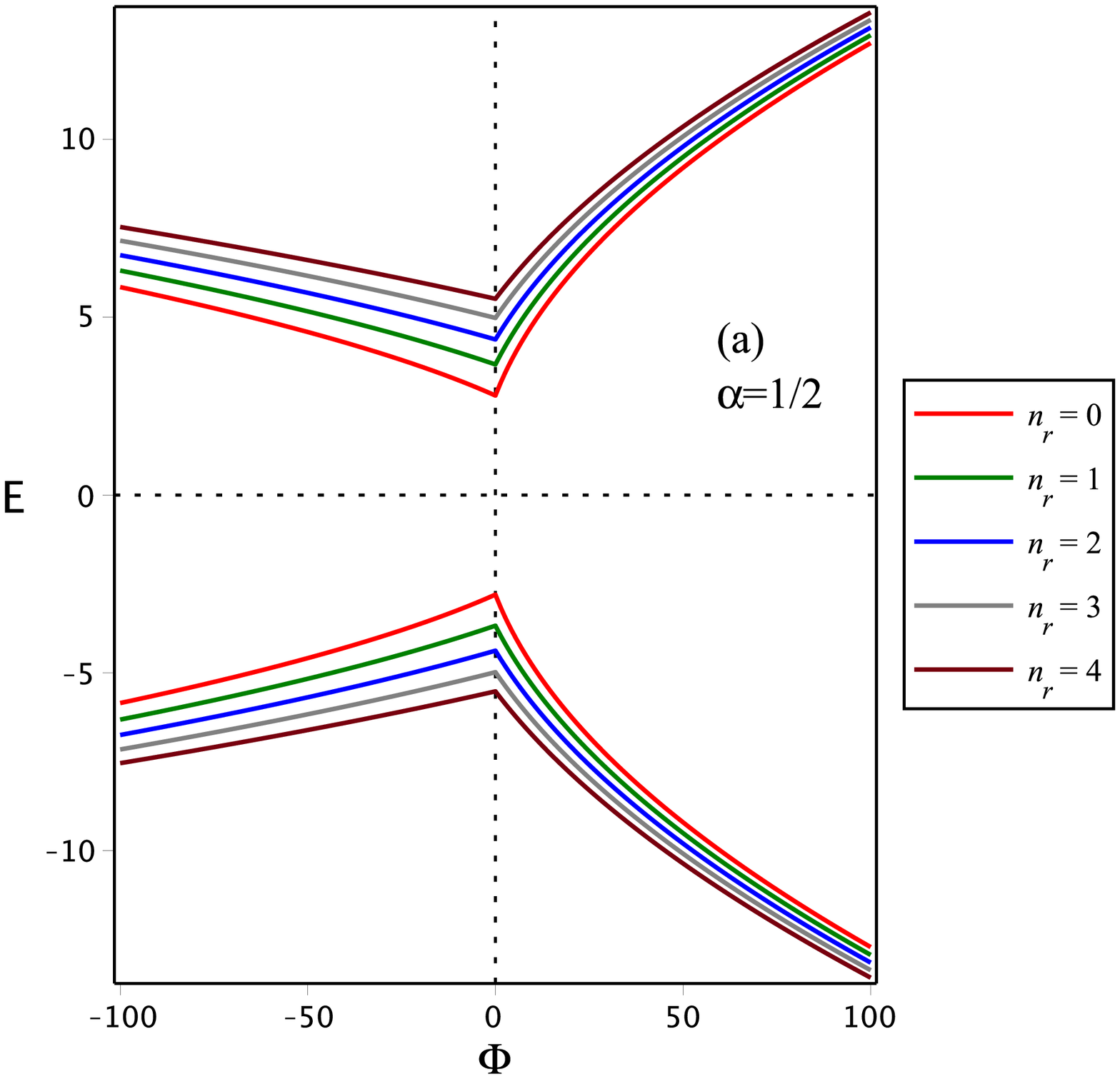}
\includegraphics[width=0.3\textwidth]{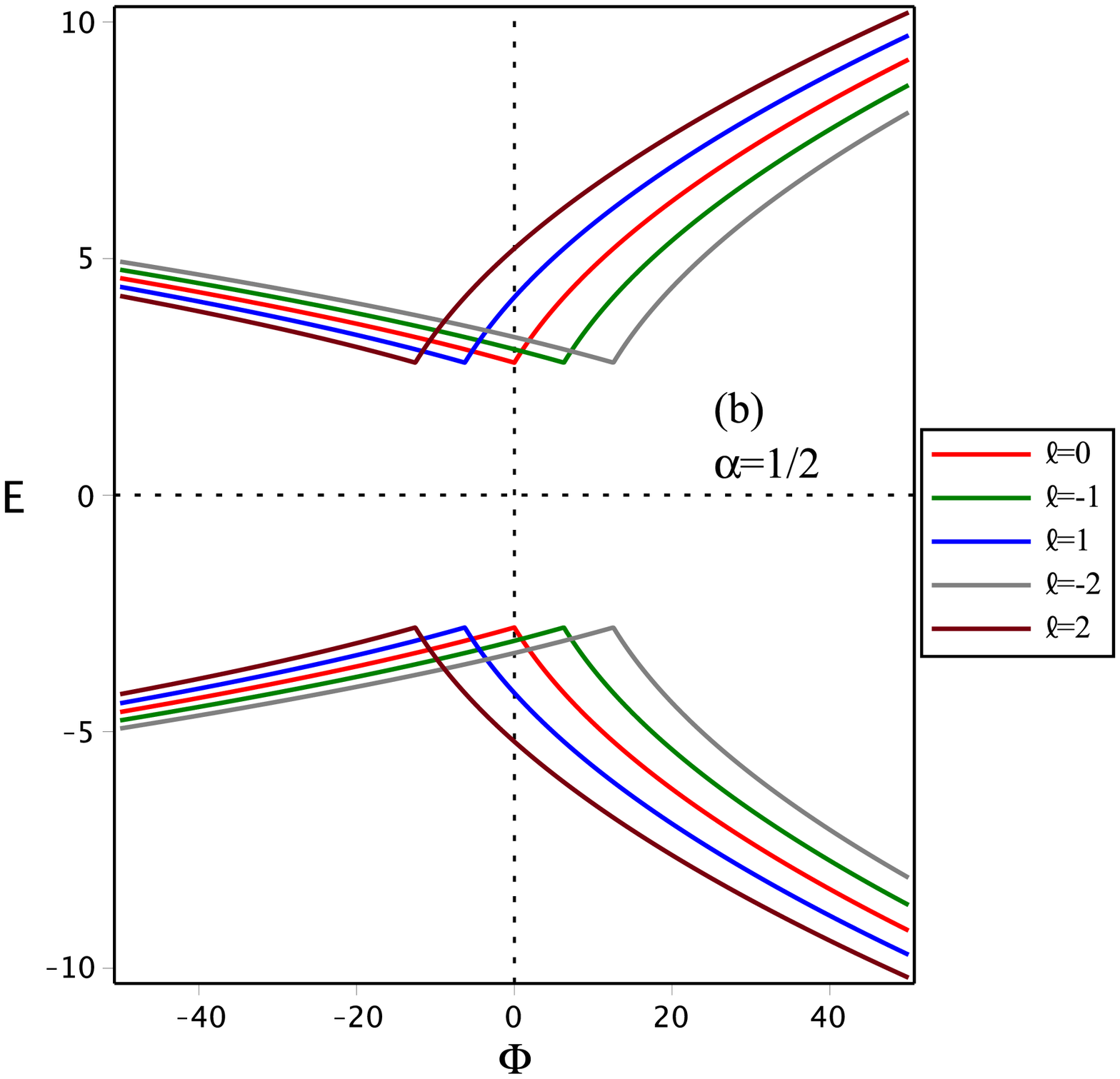}
\includegraphics[width=0.3\textwidth]{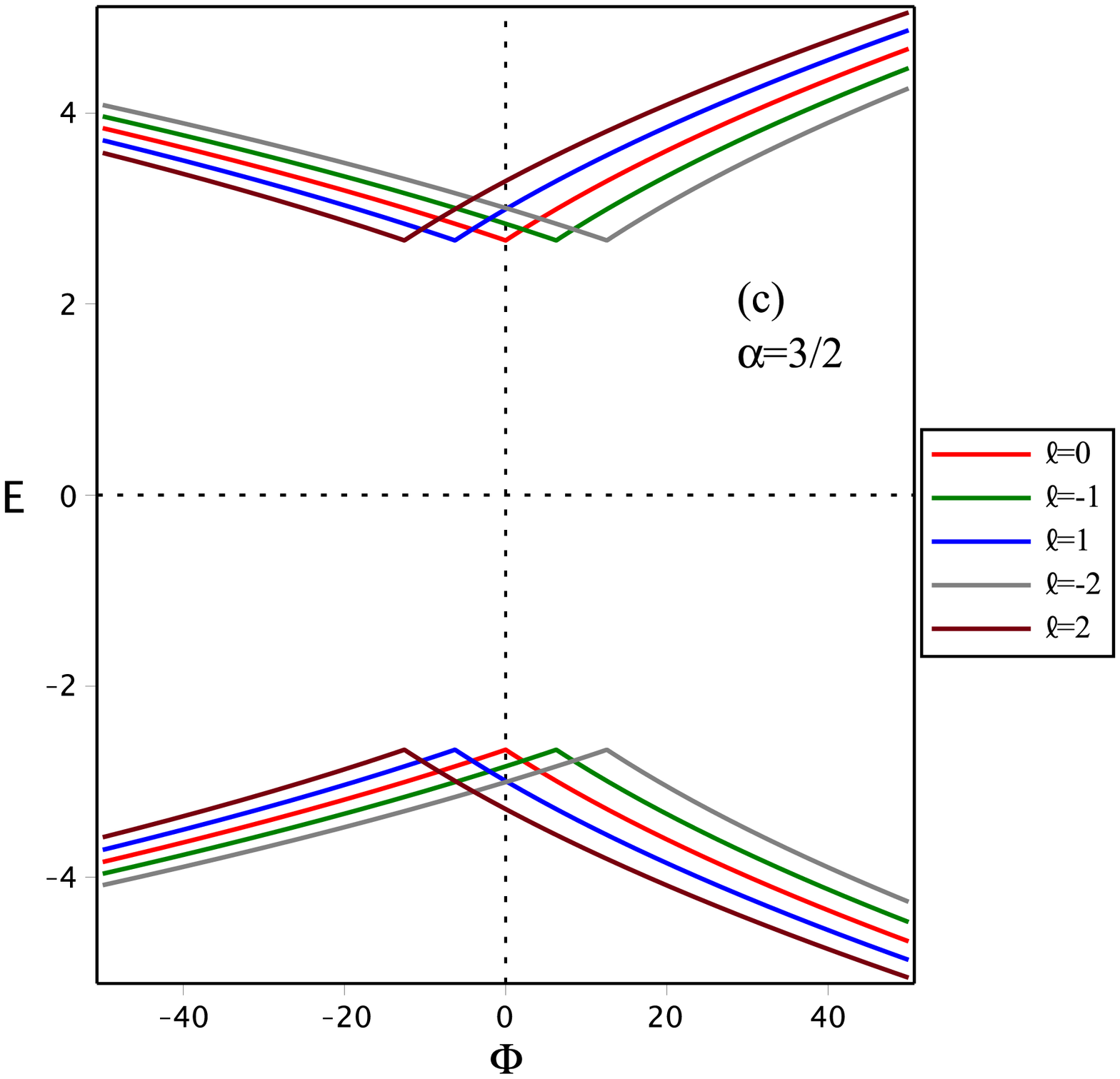}
\caption{\small 
{ For $m=k_z=Q=B_{\circ}=\eta=1$ we plot the energy levels $E_{n_r,\ell}$  so that in (a) $E_{n_r,0}$ with $\alpha=0.5<1$ (positive curvature) for $n_r=0,1,2,3,4$ at different values of $\Phi$, (b) $E_{0,\ell}$ with $\alpha=0.5$ for $\ell=0,\pm1,\pm2$ at different values of $\Phi$, and (c) $E_{0,\ell}$ with $\alpha=1.5>1$ (negative curvature) for $\ell=0,\pm1,\pm2$ at different values of $\Phi$.}}
\label{fig2}
\end{figure}%
\begin{equation}
{Q\,B_{\circ}}\,\left(\frac{2\,n_{r}}{\alpha}+\left\vert \frac{\ell }{\alpha^2}+\frac{\,\Phi\,Q}{2\,\alpha^2\,\pi}\right\vert +\frac{1}{\alpha}\right) +{Q\,B_{\circ}}\left( \frac{\ell }{\alpha^2 }+\frac{\Phi\,Q}{2\,\alpha^2\,\pi }\right).
\label{14-3}
\end{equation}%
We may see two asymptotic behaviours for (\ref{14-3}): (i) when $\alpha>>1$ we get $Q\,B_{\circ}(2\,n_r+1)/\alpha$ as the dominant term where the contribution of $\ell$  becomes insignificant (and hence for a given $n_r$ the energy  levels $E_{n_r,\ell}$  would cluster and batch up around the corresponding $E_{n_r,0}$ energy level), and (ii) as $\alpha\rightarrow0$ we get%
\begin{equation}
{Q\,B_{\circ}}\,\left(\left\vert \frac{\ell }{\alpha^2}+\frac{\,\Phi\,Q}{2\,\alpha^2\,\pi}\right\vert \right) +{Q\,B_{\circ}}\left( \frac{\ell }{\alpha^2 }+\frac{\Phi\,Q}{2\,\alpha^2\,\pi }\right).
\label{14-4}
\end{equation}%
as the dominating terms (hence states with the same $\ell=+j;\,j=0,1,2,\cdots$ value would cluster and batch up irrespective of the values of $n_r$, same for states with $\ell=-j$).

To study the effect of Aharonov-Bohm flux field $\Phi$, we plot Figure 2.  We observe that for $\Phi>0$ the energy levels are boosted up for positive energies, $E_+$, or boosted down for negative energies, $E_-$. Consequently, the energy gap widens up more rapidly than that for $\Phi<0$. This is mainly attributed to the competition between the first and second terms of (\ref{a14-2}) for $\ell=0$ in the case of Fig. 2(a) and becomes more sever in Fig. 2(b) as the magnetic quantum number $\ell$ gets more involved. This would in effect explain the energy levels shifts (i.e., to the positive $\Phi$ region for negative $\ell$ and to the negative $\Phi$ region for positive $\ell$). Hence, energy levels crossings are unavoidable in the process. Similar trend of behaviour is observed in Fig. 2(c) but with less sever shifting for $\alpha=3/2>1$ (negative curvature) than that for $\alpha=0.5<1$ (positive curvature) in 2(b).%

\section{A pseudo-confined PDM KG-oscillator in cosmic string spacetime within KKT }

A mixed exponentially growing and power-law type PDM in the form of%
\begin{equation}
m(r)=a\,\frac{e^{b\,r}}{r^\sigma}
\label{a17}
\end{equation}%
would yield%
\begin{equation}
M(r)= -\frac{\sigma^2}{16\,r^2}-\frac{1}{2}\eta\,b\,r-\frac{(\sigma -2)\,b}{8\,r}-\frac{b^2}{16}-\frac{1}{2}\sigma\,\eta.
\label{a18}
\end{equation}%
This in effect, would result that equation (\ref{a11}), with $S(r)=0$, now reads%
\begin{equation}
U^{\prime \prime }\left(r\right) +\left[\mathcal{E}-\frac{\left( 
\tilde{\beta}^{2}-1/4\right)}{r^2}-\tilde{\omega}^2\,r^2-\eta\,\tilde{a}\,r-\frac{\tilde{b}}{r}\right]\,U\left( r\right) =0,
\label{a19}
\end{equation}%
and%
\begin{equation}
\mathcal{E}=E^2-Q^2-{k_z}^2-m^2-2\,\omega\,\tilde{\gamma}-\frac{1}{2}(\sigma+4)\,\eta-\frac{b^2}{16},
\label{a19-1}
\end{equation}
where%
\begin{equation}
\tilde{a}=b/2,\,\,\tilde{b}=b(\sigma-2)/8, \,\,\tilde{\omega}^2=\omega^2+\eta^2,\,\,\tilde{\beta}^2=\tilde{\gamma}^2+\sigma^2/16.
\label{a19-2}
\end{equation}%
Obviously, such PDM has introduced its own Cornell-type (i.e., $\eta\,\tilde{a}\,r+{\tilde{b}}/{r}$) confinement as a PDM-byproduct. Hence, the notion of pseudo-confined PDM KG-oscillator is introduced.  Moreover, this differential equation admits a solution in the form of biconfluent Heun function given by%
\begin{equation}
U(r)=\mathcal{N}\, r^{\vert\tilde{\beta}\vert+1/2}\,exp\left(-\frac{1}{2}\tilde{\omega}\,r^2+\frac{\tilde{a}\,\eta\,r}{2\tilde{\omega}}\right)\,H_{B}\left(2\vert\tilde{\beta}\vert,\,\frac{\tilde{a}\eta}{\tilde{\omega}^{3/2}},\,\frac{\tilde{a}^2\eta^2+4\mathcal{E}\tilde{\omega}^2}{4\,\tilde{\omega}^3},\,\frac{2\,\tilde{b}}{\sqrt{\tilde{\omega}}},\,\sqrt{\tilde{\omega}}\,r\right).
\label{a20}
\end{equation}%
To retrieve the two-dimensional oscillator eigenvalues $\tilde{\lambda}=\tilde{\mathcal{E}}$, one may set $\tilde{a}=\tilde{b}=0$ in (\ref{a19}) to obtain%
\begin{equation}
U(r)=\mathcal{N}\, r^{\vert\tilde{\beta}\vert+1/2}\,exp\left(-\frac{1}{2}\tilde{\omega}\,r^2\right)\,H_{B}\left(2\vert\tilde{\beta}\vert,\,0,\,\frac{\tilde{\mathcal{E}}}{\tilde{\omega}},0,\,\sqrt{\tilde{\omega}}\,r\right).
\label{a21}
\end{equation}%
In this case, we again use the Kummer relation \cite{Ron 1995}%
\begin{equation}
H_{B}\left(2\vert\tilde{\beta}\vert,\,0,\frac{\tilde{\mathcal{E}}}{\tilde{\omega}},0,\,\sqrt{\tilde{\omega}}\,r\right)=\,_{1}F_{1}\left(\left[\frac{2\,\vert\tilde{\beta}\vert\,\tilde{\omega}-\tilde{\mathcal{E}}+2\,\tilde{\omega}^2}{4\,\tilde{\omega}}\right],\,\left[1+\vert\tilde{\beta}\vert\right],\,\tilde{\omega}\,r\right)
\label{a22}
\end{equation}%
to imply that%
\begin{equation}
\frac{2\,\vert\tilde{\beta}\vert\,\tilde{\omega}-\tilde{\mathcal{E}}+2\,\tilde{\omega}^2}{4\,\tilde{\omega}}=\,-n_r\Rightarrow\, \tilde{\mathcal{E}}=2\,\tilde{\omega}\left(2\,n_r+\vert\tilde{\beta}\vert+1\right)
\label{a23}
\end{equation}%
This would give us the relation between $\tilde{\alpha}$ and $\gamma$ in the biconfluent Heun function $H_{B}(\tilde{\alpha},\beta,\gamma,\delta,z)$ as $\gamma=2(2n_r+1)+\tilde{\alpha}$, so that a finite biconfluent Heun polynomial of degree $2n_r\geq0$ (i.e., of degree $n_r\geq0$ ) is obtained. Consequently, for our biconfluent Heun polynomial in (\ref{a20}) we get%
\begin{equation}
\frac{\tilde{a}^2\eta^2+4\mathcal{E}\tilde{\omega}^2}{4\,\tilde{\omega}^3}=2(2n_r+1)+2\vert\tilde{\beta}\vert\Rightarrow\mathcal{E}=2\,\tilde{\omega}\left(2\,n_r+\vert\tilde{\beta}\vert+1\right)-\frac{\tilde{a}^2\eta^2}{4\,\tilde{\omega}^2}.
\label{a24}
\end{equation}%
Which would result, along with (\ref{a19-1}) and (\ref{a19-2}), the energy eigenvalues as%
\begin{equation}
E^2=2\,\tilde{\omega}\left(2\,n_r+\vert\tilde{\beta}\vert+1\right)+k_{z}^2+Q^2+m^2+2\,{\omega}\,\tilde{\gamma}+\frac{b^2}{16}+\frac{\sigma+4}{2}\,\eta
-\frac{\tilde{a}^2\eta^2}{4\,\tilde{\omega}^2}
\label{a25}
\end{equation}%
One should notice that this result collapses into that in (\ref{a14-2}) and (\ref{a15}) for $m(r)=1$ for $\sigma=0=b$ and $a=1$ of (\ref{a17}) as a natural tendency of the solution of the more general problem at hand. Therefore, the parametric settings adopted above, for our biconfluent Heun polynomial of (\ref{a20}) are reasonable/consistent, and mathematically and/or quantum mechanically acceptable settings.%
\begin{figure}[!ht]  
\centering
\includegraphics[width=0.3\textwidth]{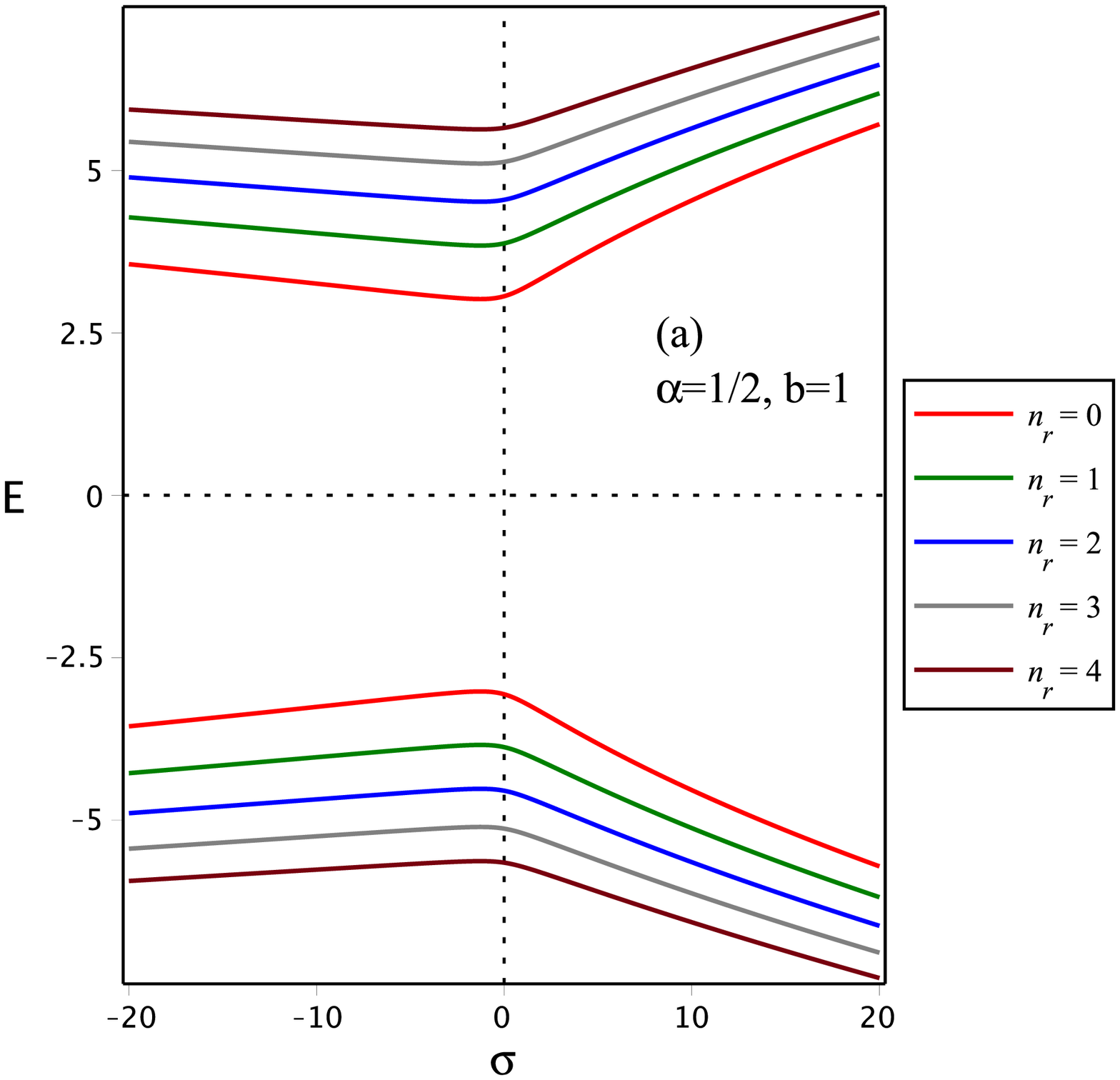}
\includegraphics[width=0.3\textwidth]{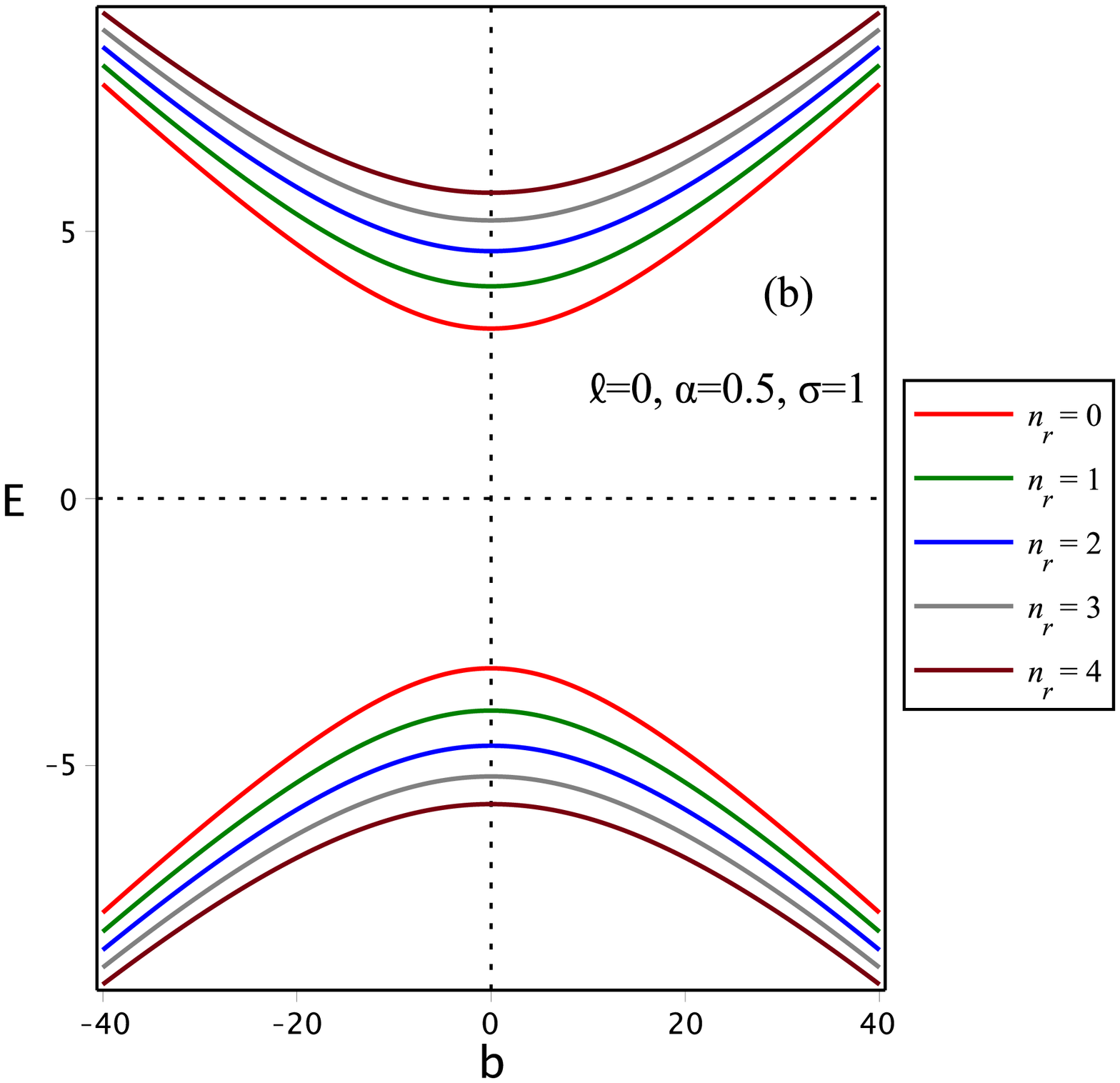}
\includegraphics[width=0.3\textwidth]{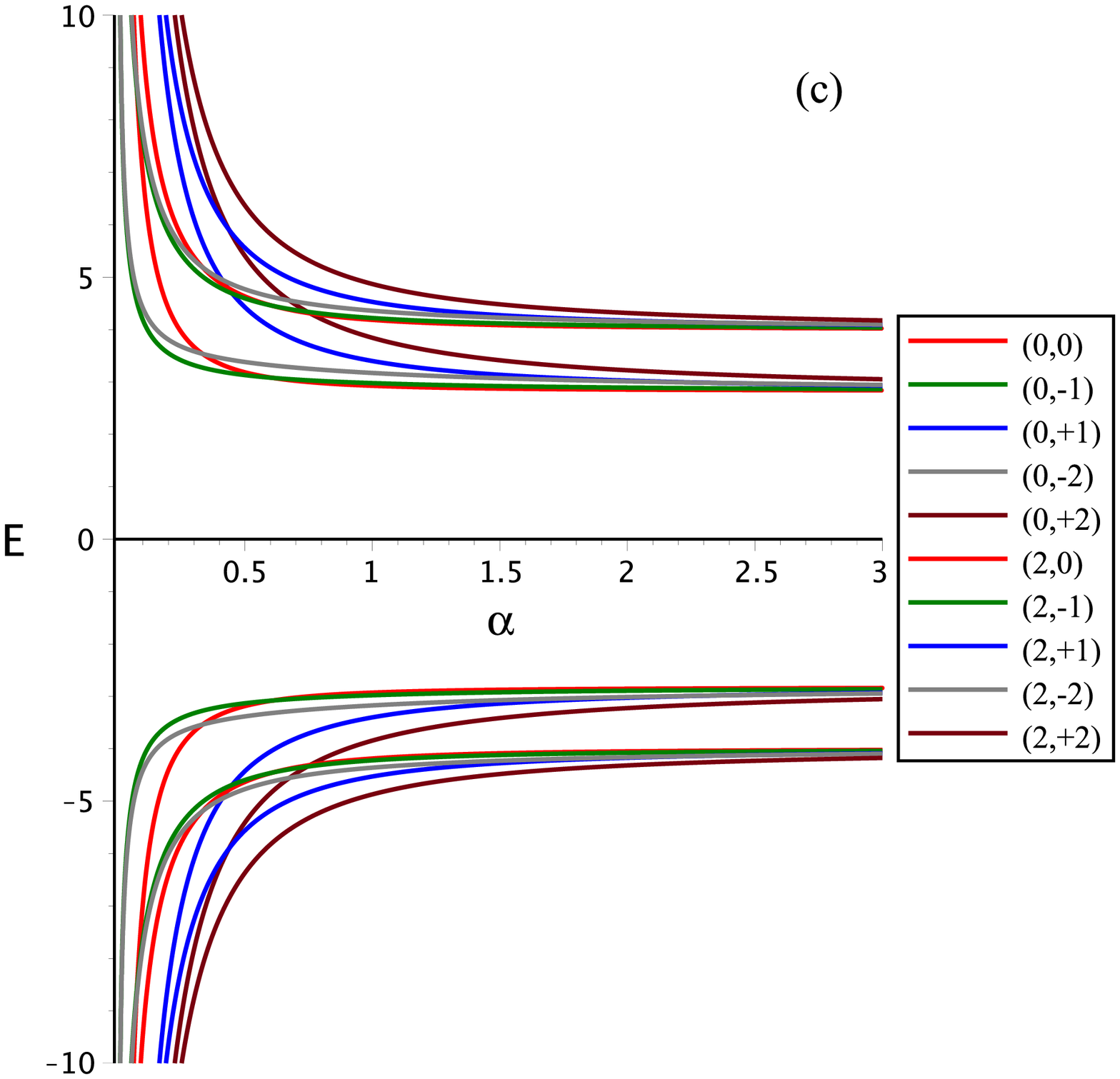}
\caption{\small 
{ For $m=k_z=Q=\Phi=B_{\circ}=\eta=1$ we show the effect of the PDM in (\ref{a17}) we plot the energy levels $E_{n_r,\ell}$  so that in (a) $E_{n_r,0}$ with $\alpha=0.5$ and $b=1$ for $n_r=0,1,2,3,4$ at different values of the PDM parameter $\sigma$, (b) $E_{n_r,0}$ with $\alpha=0.5$ and $\sigma=1$ for $n_r=0,1,2,3,4$ at different values of $b$, and $\ell=0,\pm1,\pm2$. For $m=k_z=Q=B_{\circ}=\eta=1$,  $\Phi=0$ we plot in (c) $E_{n_r,\ell}$ with $\sigma=b=1$ for $n_r=0,2$ and $\ell=0,\pm1,\pm2$ at different values of $\alpha$.}}
\label{fig3}
\end{figure}%

In Fig.3, we show the energy levels (\ref{a25}) corresponding to the PDM of (\ref{a17}). Where we observe energy levels clustering as the PDM parameter $|\sigma|$ grows up from zero (for a fixed value of the PDM parameter $b$) in Fig. 3(a), and as the PDM parameter $|b|$ grows up from zero (for a fixed value of $\sigma$) in Fig. 3(b). In Fig. 3(c), similar trends of  clustering, batching-up, and energy levels crossings (as in 1(b) and 1(c)) are observed.

\section{A confined PDM KG-oscillator-III in cosmic string spacetime within KKT}
Let us now consider the PDM KG-oscillator-III, with%
\begin{equation}
m(r)=\tilde{A}\,e^{2\,\Omega\,r^2}
\label{a25-1}
\end{equation}%
of (\ref{a12-0}), subjected to a Cornell-type confinement $S(r)=A\,r+B/r$. In this case, equation (\ref{a12}) yields $M(r)=-(\Omega^2\,r^2+2\,\Omega)$ and equation (\ref{a11}) reads%
\begin{equation}
U^{\prime \prime }\left(r\right) +\left[\tilde{\lambda}_1-\frac{\left( 
{\tilde{\gamma}_1}^{2}-1/4\right)}{r^2}-{\tilde{\omega}_1}^2\,r^2-2\,m\,A\,r-\frac{2\,B}{r} \right]\,U\left( r\right) =0,
\label{a26}
\end{equation}%
where $\tilde{\lambda}_1=\lambda-2\,(\eta+\Omega)-2AB$,  ${\tilde{\omega}_1}^2=\omega^2+(\eta^2+\Omega^2)+A^2$, and ${\tilde{\gamma}_1}^{2}=\tilde{\gamma}^2+B^2$.  This differential equation admits a solution in the form of biconfluent Heun function given by%
\begin{equation}
U(r)=\mathcal{N}\, r^{\vert{\tilde{\gamma}_1}\vert+1/2}\,exp\left(-\frac{1}{2}{\tilde{\omega}_1}\,r^2+\frac{m\,A\,r}{{\tilde{\omega}_1}}\right)\,H_{B}\left(2\vert{\tilde{\gamma}_1}\vert,\,\frac{2\,m\,A}{{\tilde{\omega}_1}^{3/2}},\,\frac{m^2A^2+\tilde{\lambda}_1\,{\tilde{\omega}_1}^2}{{\tilde{\omega}_1}^3},\,\frac{4\,m\,B}{\sqrt{{\tilde{\omega}_1}}},\,\sqrt{{\tilde{\omega}_1}}\,r\right).
\label{a27}
\end{equation}%
In the above section, we have readily asserted that the relation between $\tilde{\alpha}$ and $\gamma$ in the biconfluent Heun function $H_{B}(\tilde{\alpha},\beta,\gamma,\delta,z)$ is given by $\gamma=2(2n_r+1)+\tilde{\alpha}$. This would result in a biconfluent Heun polynomial of degree $2n_r\geq0$ (i.e., of degree $n_r\geq0$ ) to secure the finiteness of the wave function. Consequently, for our biconfluent Heun polynomial in (\ref{a27}) we get%
\begin{equation}
\frac{m^2A^2+\tilde{\lambda}_1\,{\tilde{\omega}_1}^2}{{\tilde{\omega}_1}^3}=2(2n_r+1)+2\vert{\tilde{\gamma}_1}\vert\Rightarrow\tilde{\lambda}_1=2\,{\tilde{\omega}_1}\left(2\,n_r+\vert{\tilde{\gamma}_1}\vert+1\right)-\frac{m^2A^2}{{\tilde{\omega}_1}^2}.
\label{a28}
\end{equation}%
Which would result%
\begin{equation}
E^2=2\,{\tilde{\omega}_1}\left(2\,n_r+\vert{\tilde{\gamma}_1}\vert+1\right)+2\,{\omega}\,\tilde{\gamma}+k_{z}^2+Q^2+m^2+2\,[\eta+\Omega]+2\,A\,B-\frac{m^2A^2}{{\tilde{\omega}_1}^2}.
\label{29}
\end{equation}%
\begin{figure}[!ht]   
\centering
\includegraphics[width=0.3\textwidth]{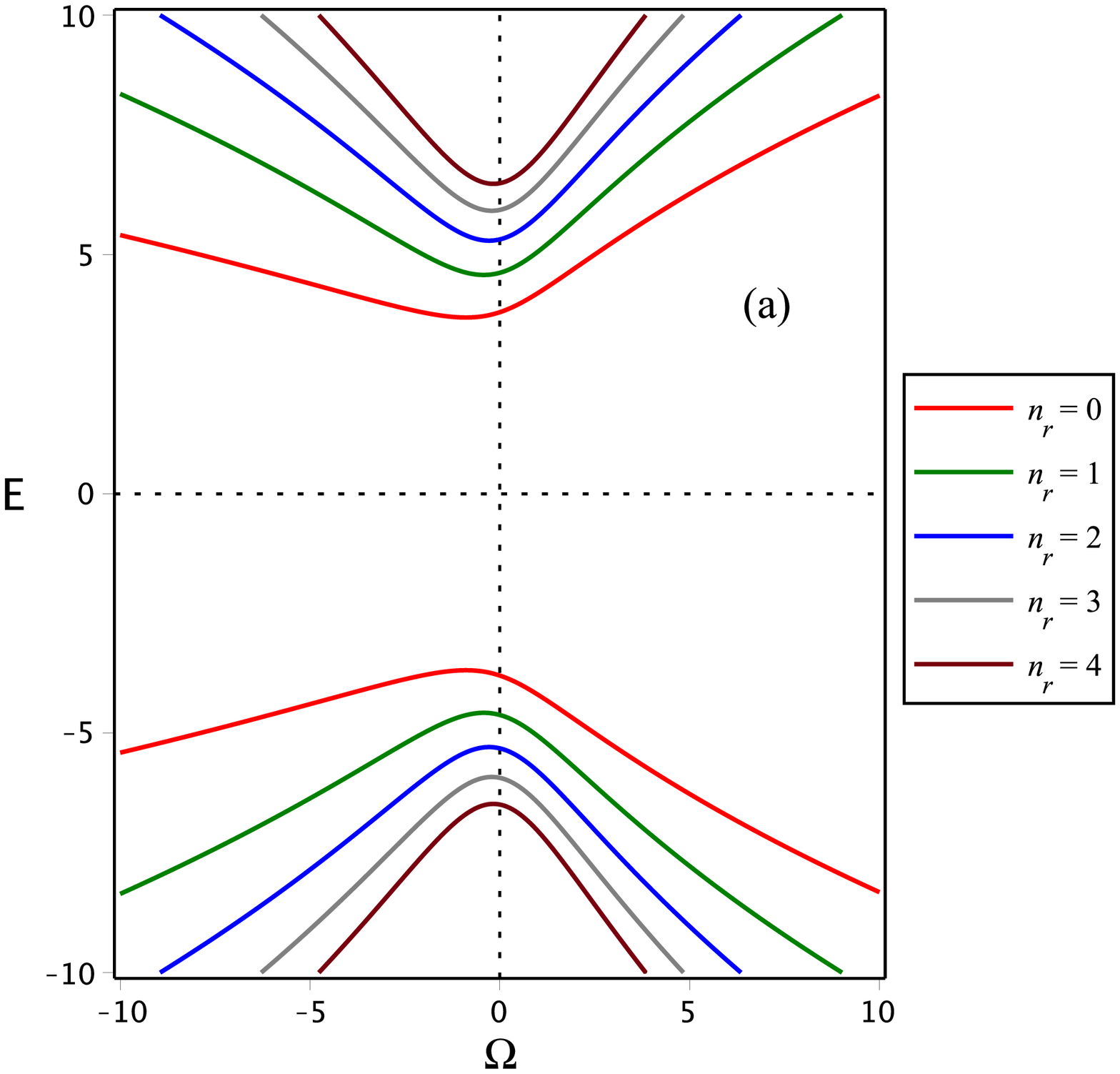}
\includegraphics[width=0.3\textwidth]{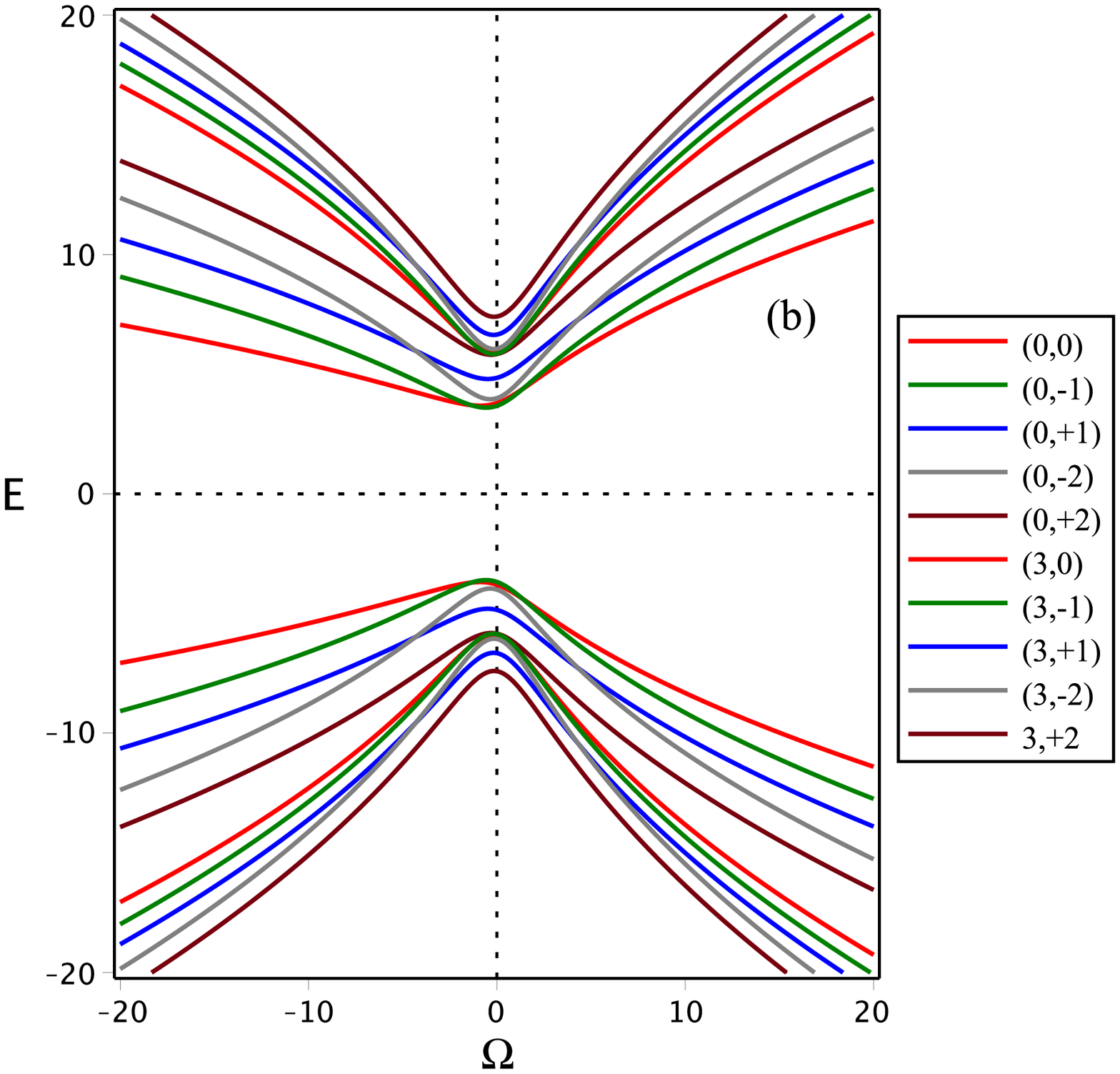}
\includegraphics[width=0.3\textwidth]{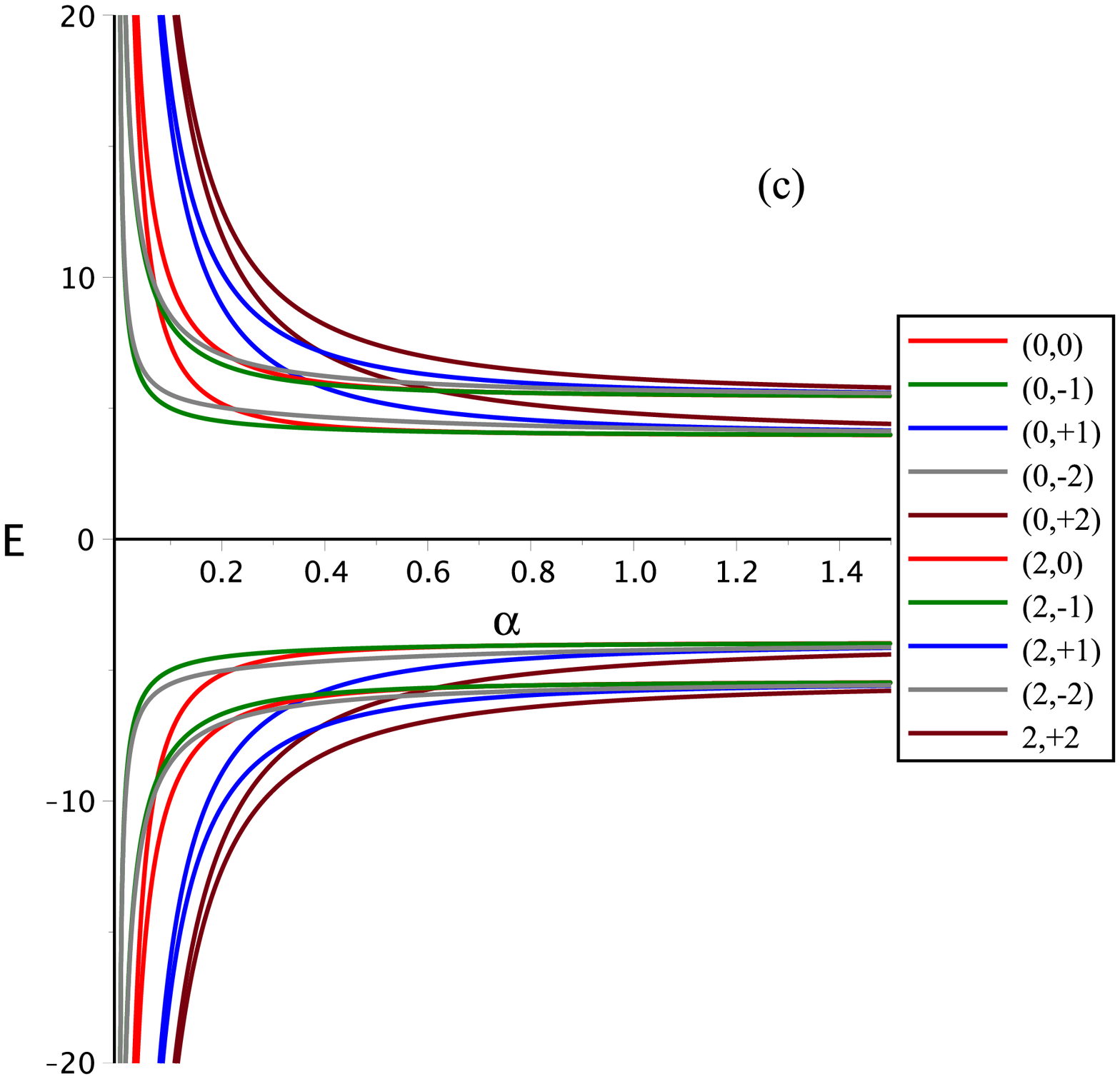}
\caption{\small 
{ For $m=k_z=Q=\Phi=B_{\circ}=\eta=A=B=1$ and $\alpha=0.5$ we show the effect of the PDM parameter $\Omega$ in (\ref{a25-1}) on the energy levels $E_{n_r,\ell}$. We plot in (a) $E_{n_r,0}$ states for $\ell=0$ and and $n_r=0,1,2,3,4$ at different values of the PDM parameter $\Omega$, (b) $E_{n_r,\ell}$ for $\ell=0,\pm1,\pm2$ and $n_r=0,3$ at different values of $\Omega$, and (c) $E_{n_r,\ell}$ with $\Omega=1$ for $n_r=0,2$ and $\ell=0,\pm1,\pm2$ at different values of $\alpha$. }}
\label{fig5}
\end{figure}%
In Figure 4, we show the effect of the PDM parameter $\Omega$ on the energy levels, which are readily quasi-Landau levels. In Fig. 4(a) we see that the energy gap increases more rapidly for $\Omega>0$ than that for $\Omega<0$. This is because of the contribution of the sixth term in (\ref{29}). In Fig 4(b), we observe energy levels crossing because of the irrational magnetic quantum number $\tilde{\gamma}$ involved in the first and second terms of (\ref{29}), where the PDM parameter $\Omega$  is only indulged in the first term as $\Omega^2$ (consequently, positive or negative $\Omega$ have the same effect). In Fig. 4(c), we again observe the same trend of behaviour as that readily discussed for Fig. 1(b).%
\section{$\mathcal{PT}$-symmetric PDM KG-particle in cosmic string spacetime within KKT}
The first interest in non-Hermitian Hamiltonians date back to Caliceti et al  \cite{Caliceti 1980} which offered the first rigorous explanation why the spectrum in non-Hermitian model may be real and discrete. This has initiated extensive discussions (cf., e.g., \cite{Buslaev 1993}) which eventually resulted in the proposal of $\mathcal{PT}$-symmetric quantum mechanics of Bender and Boettcher \cite{Bender 1998}. Where the key idea lies in the existence of the real spectrum need not necessarily be attributed to the Hermiticity of the Hamiltonian. The textbook Hermiticity assumption $H=H^{\dagger}$ is replaced by the mere $\mathcal{PT}$-symmetry $H=H^{\ddagger}=\mathcal{PT}H\mathcal{PT}$ (for more details on this issue the reader may refer to \cite{Znojil Levai 2000, Mustafa Znojil 2002} and references cited therein). Hereby, $\mathcal{P}$ denotes space reflection: $\mathcal{PT}x\,\mathcal{PT}\rightarrow-x$, $\mathcal{T}$ mimics the time-reversal: $\mathcal{PT}i\,\mathcal{PT}\rightarrow-i$.  Under such non-Hermitian settings, we use a $\mathcal{PT}$-symmetric PDM so that%
\begin{equation}
m(r)=a\,e^{4\, i\tilde{A}\,r}.
\label{PT PDM}
\end{equation}%
This would, in turn, imply that equation (\ref{a12}), with $\eta=0$ now reads%
\begin{equation}
M(r)={\tilde{A}}^2-\frac{i\,\tilde{A}}{r}.
\label{PT M(r)}
\end{equation}%
Under such settings and  equation (\ref{a11}) yields%
\begin{figure}[!ht]   
\centering
\includegraphics[width=0.3\textwidth]{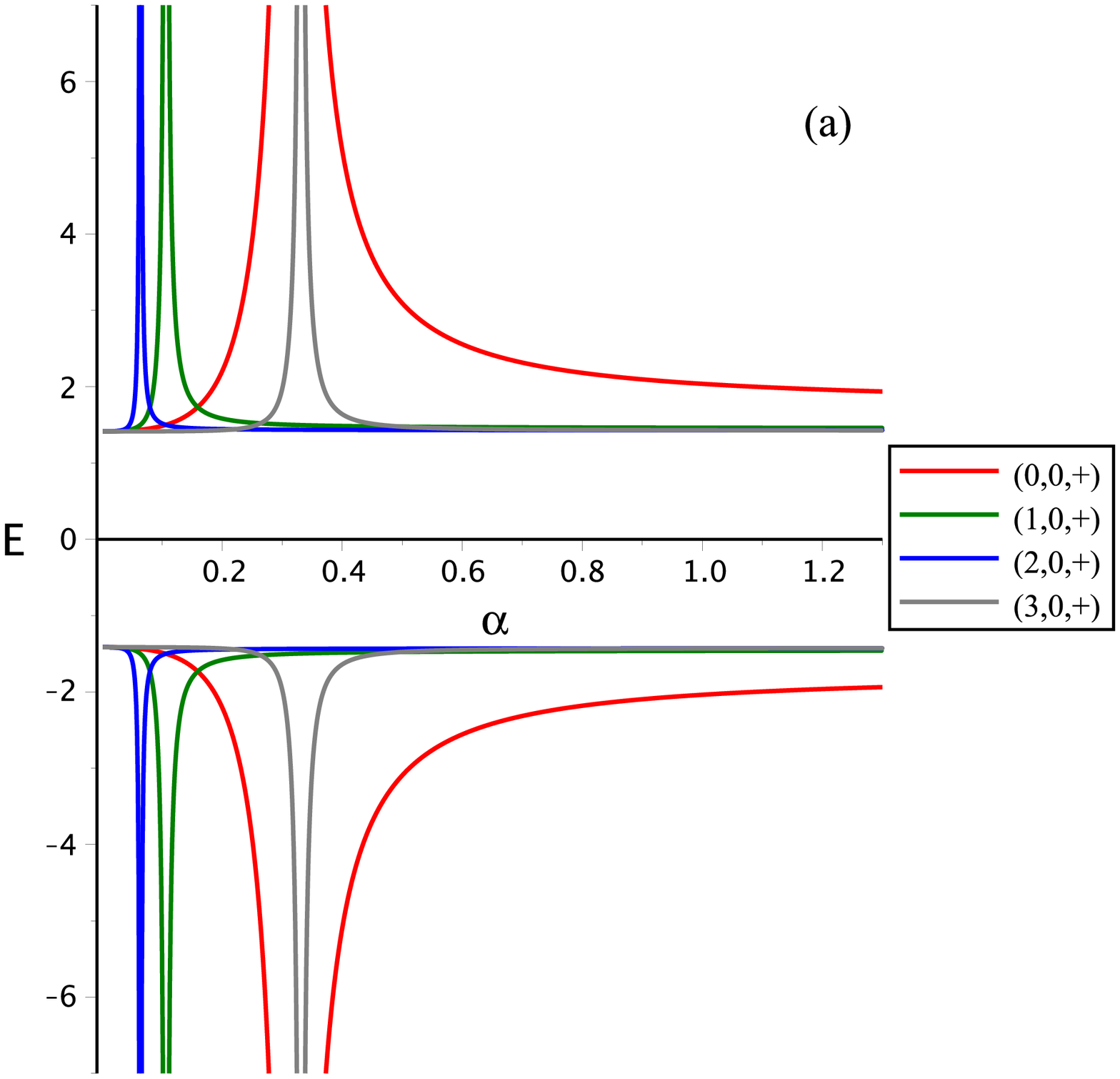}
\includegraphics[width=0.3\textwidth]{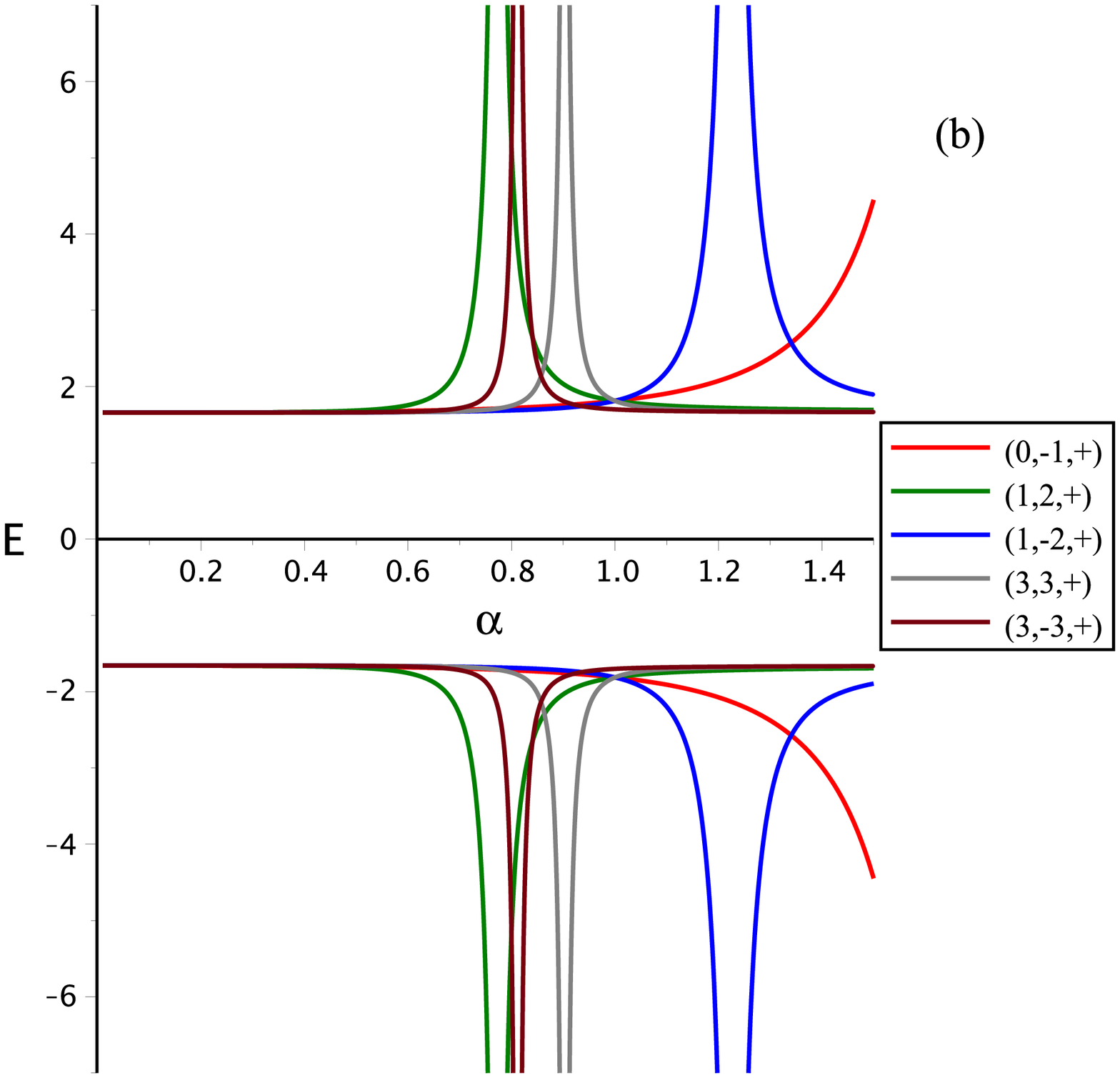}
\includegraphics[width=0.3\textwidth]{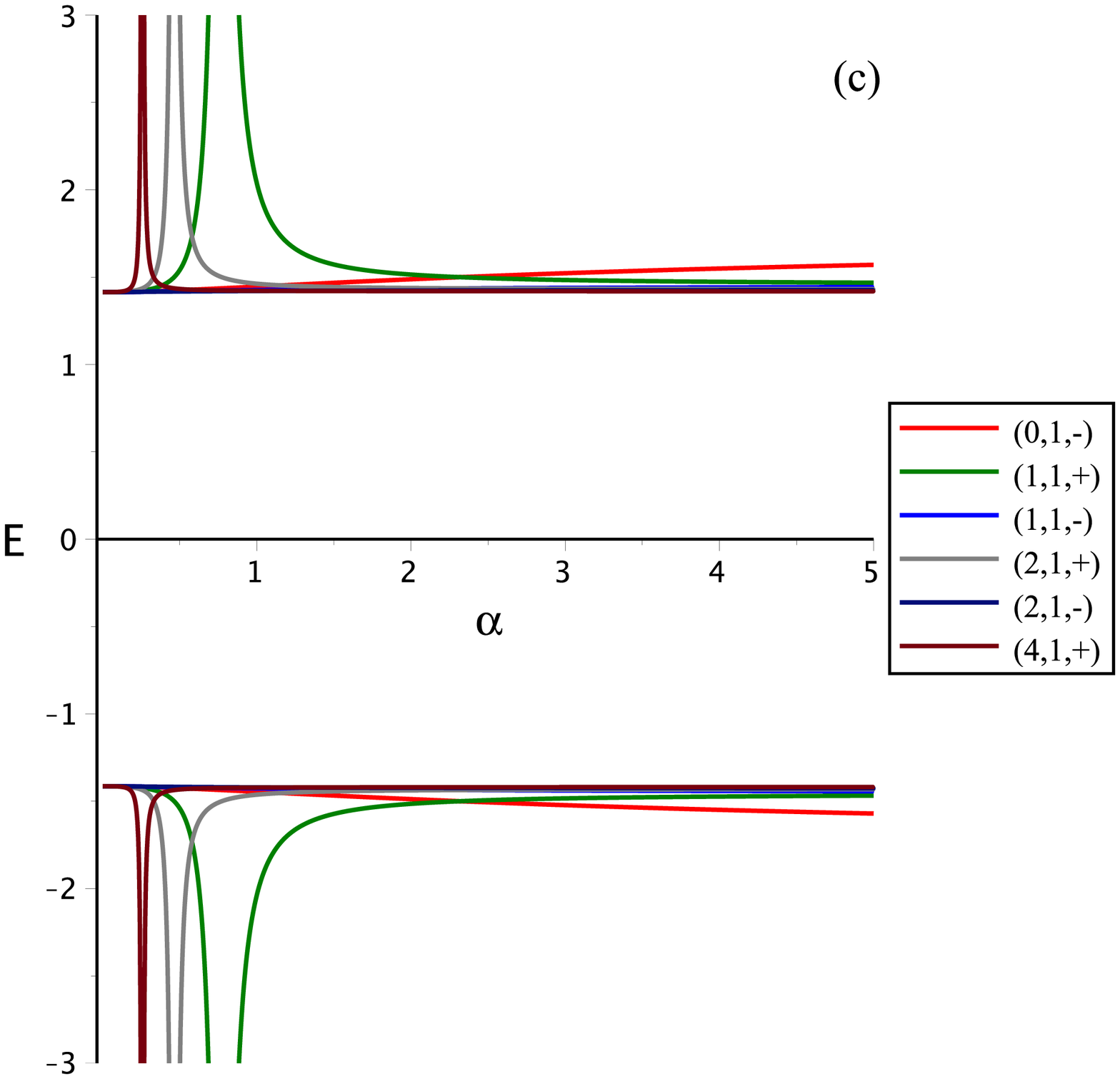}
\caption{\small 
{ For $m=k_z=Q=\Phi=\tilde{A}=1$ we plot $E_{n_r,\,\ell,\,q}$ against $\alpha$ (a) for $n_r=0,1,2,3$, $\ell=0$, and with even quasi-parity $q=+1$, (b) for $n_r=0,1,3,\,\ell=-1,\pm2,\pm3$ , and with even quasi-parity $q=+1$, and (c) for $n_r=0,1,2,4,\,\ell=1$, and $q=\pm1$. }}
\label{fig5}
\end{figure}%
\begin{equation}
U^{\prime \prime }\left(r\right) +\left[\tilde{\mathcal{E}}-\frac{\left( 
\tilde{\gamma}^{2}-1/4\right)}{r^2}-\frac{i\,\tilde{A}}{r} \right]\,U\left( r\right) =0,
\label{PT eq}
\end{equation}%
where, $\tilde{\mathcal{E}}=E^{2}-Q^{2}-k_{z}^{2}-m^{2}+{\tilde{A}}^2$, $\tilde{\gamma}^{2}=\left( \frac{\ell }{\alpha }+\frac{ \Phi Q}{2\,\alpha\,\pi  }\right) ^{2}$ and our PDM KG-Coulombic particle is only under Aharonov-Bohm flux field and no magnetic field is involved in this process (i.,e.,  $B_{\circ }=0$). This Schr\"{o}dinger-like Hamiltonian (i.e., $\hat{H}U(r)=\tilde{\mathcal{E}}U(r)$) is $\mathcal{PT}$- symmetric by structure and admits real eigenvalues $\tilde{\mathcal{E}}\in\mathbb{R}$.  In fact, this equation (\ref{PT eq}) represents the so called $\mathcal{PT}$-symmetric Schr\"{o}dinger-Coulomb problem.  It has been thoroughly discussed by  Znojil and L\'{e}vai \cite{Znojil Levai 2000} and later on by Mustafa and Znojil \cite{Mustafa Znojil 2002} using the so called $\mathcal{PT}$-symmetric pseudo-perturbation shifted-$\ell$ expansion quasi-perturbation recipe. Where, it has been shown, in \cite{Znojil Levai 2000,Mustafa Znojil 2002},  that if the central-like repulsive/attractive core, in \cite{Mustafa Znojil 2002}, is replaced through the transformation $\ell_d (\ell_d+1)\rightarrow \tilde{\gamma}^{2}-1/4$ , i.e. $\ell_d=q\,\vert\tilde{\gamma}\vert-1/2$, with $q=\pm1$ denoting \textit{quasi-parity}, then the reported exact eigenvalues $\tilde{\mathcal{E}}$ would read%
 \begin{equation}
 \tilde{\mathcal{E}}=\frac{{\tilde{A}}^2}{4\,(n_r-\ell_d)^2}\Rightarrow \tilde{\mathcal{E}}=\frac{{\tilde{A}}^2}{(2\,n_r-2\,q\,\vert\tilde{\gamma}\vert+1)^2}.
 \label{E Coulomb0}
 \end{equation}%
 At this point, one should notice that the degeneracy associated with the regular Hermitian Coulomb  eigenvalues $\tilde{\mathcal{E}}=-{\tilde{A}}^2/(2\,n^2);\,n=n_r+\vert\ell\vert+1/2=1,\,2,\cdots$, is now lifted upon the complexification of, say, the dielectric constant embedded in $\tilde{A}$. Hence,  the energies for our $\mathcal{PT}$-symmetric KG-Coulombic particle in cosmic string spacetime within KKT are given by%
 \begin{equation}
 E^2=\frac{{\tilde{A}}^2}{(2\,n_r-2\,q\,\vert\tilde{\gamma}\vert+1)^2}-{\tilde{A}}^2+Q^{2}+k_{z}^{2}+m^{2},
 \label{PT energy}
 \end{equation}%
 and  the radial wave functions \cite{Znojil Levai 2000} are given by  %
 \begin{equation}
U\left( r\right) \sim r^{\frac{1}{2}-q\,\left\vert \tilde{\gamma}\right\vert }\exp
\left( i\,{\tilde{A}}^2\,r\right) L_{n_{r}}^{-2\,q\,\left\vert \tilde{\gamma}\right\vert }\left( -2\,i\,{\tilde{A}}^2\,r\right) \Longrightarrow
R\left( r\right) \sim r^{-q\,\left\vert \tilde{\gamma}\right\vert }\exp
\left( i\,{\tilde{A}}^2\,r\right) L_{n_{r}}^{-2\,q\,\left\vert \tilde{\gamma}\right\vert }\left( -2\,i\,{\tilde{A}}^2\,r\right),
\label{R(r) PT}
\end{equation}%

Moreover, the energies in (\ref{PT energy}) suggest the phenomenon of \textit{flown away states} at $n_r=q\,\vert\tilde{\gamma}\vert-1/2\geq0$ (i.e., for each $n_r$-state there is an $\ell_d=q\,\vert\tilde{\gamma}\vert-1/2$ state to fly away). Under such condition, only states with even quasi-parity, $q=+1$, are destined to fly away at some specific values of the curvature parameter $\alpha$.  For more details on the issue of \textit{flown-away} states and thorough discussion on the $\mathcal{PT}$-symmetric Coulomb problem the reader is advised to refer to  \cite{Znojil Levai 2000,Mustafa Znojil 2002} and related references cited therein.

In Figure 5, we show the energy levels  $E_{n_r,\,\ell,\,q}$ against $\alpha$. For states with even quasi-parity, we observe energy levels crossings (in 5(a) for different $n_r$ and $\ell=0$, and in 5(b) for different $n_r$ and different $\ell$.  Whereas, in Fig.5(c) we observe energy levels crossings between states with even and odd quasi-parities. That is, a state with odd quasi-parity $n_r'$ crosses with a state with even parity $n_r$ when the relation $\alpha\,(n_r-n_r')=2\,\vert\ell+\frac{\Phi\,Q}{2\,\pi}\vert$ is satisfied. %

\section{Concluding remarks}
In the current methodical proposal, we have introduced the effective PDM setting for the KG-oscillators (in particular and KG-particles in general) in cosmic string spacetime in magnetic and Aharonov-Bohm flux fields within the Kaluza-Klein theory. The effective PDM is introduced as a defect/deformation in the momentum operator analogous to effective PDM in Schr\"{o}dinger theory. We have reported, in section 2, two new PDM KG-oscillators among four feasible KG-oscillators in the cosmic string spacetime within KKT. Therein, we have shown that the KG-oscillator-I by Mirza's et al \cite{Mirza 2004} turns out to be nothings but a byproduct of the PDM setting in KG-oscillator-III of (\ref{a12-0}). Effectively, the topological defect in the momentum operator suggested by Mirza et al. may very well be explained as an effective topological defect introduced by effective PDM settings.  Moreover, we have observed that square the energies in (\ref{a14-2}) (i.e., $E^2$) resembles the Landau-type energies-squared with an irrational magnetic quantum number $\tilde{\gamma}$ that indulges within the Aharonov-Bohm flux field effect. We may argue, therefore, that the additional compact dimension of the Kaluza-Klein theory offers quasi-Landau energy levels (e.g.,  \cite{Geusa 2001,Bezerra 2019,Zeinab 2020}). Hereby, the behaviour of the energy  levels $E_{n_r,\ell}$ is found to follow two different trends of clustering and batching up for the curvature parameter $\alpha\approx 0$ and $\alpha\gg 1$. This would necessarily suggest that energy levels crossings are unavoidable in the process (documented in Fig.s 1(b), 1(c), 2(b), 2(c), 3(c), 4(b), and 4(c)). Whereas, at $\alpha=1$ (i.e., Minkowski flat spacetime) the energy levels maintain their regular quasi-Landau-type structure. 

Next, we have discussed, in section 4, a mixed power-law and exponential type  PDM model that resulted in a pseudo-confined PDM KG-oscillator in cosmic string spacetime within KKT. The effective confining potential is introduced as a byproduct of the PDM settings of the KG-oscillator. In this case, the PDM KG-oscillators are shown to be confined in their own PDM manifested effective interaction potential force field, hence the notion \textit{"pseudo-confined PDM KG-oscillators"} is unavoidably inviting.  We have considered, in section 5,  a confined PDM KG-oscillator-III with the PDM (\ref{a25-1}) which could basically replace Mirza et al's recipe \cite{Mirza 2004} with $\Omega=\eta$ in (\ref{a26}). Finally, we have reported on a non-Hermitian $\mathcal{PT}$-symmetric PDM-Coulombic KG-particle model in cosmic string spacetime within KKT. Some unusual quantum mechanical properties such as \textit{flown away states} are observed. The detailed discussions on this issue may be found in (cf., e.g.,  \cite{Bender 1998,Znojil Levai 2000,Mustafa Znojil 2002}). The current methodical proposal should, therefore, be considered as a new generalization where its applicability covers not only PDM KG-oscillators (documented in sections 3, 4, and 5) but also PDM KG-particles (documented in section 6) in cosmic string spacetime or in the geometric theory of topological defects in condensed matter. To the best of our knowledge, the current methodical proposal did not appear elsewhere.

\end{document}